\documentclass[12pt,a4paper]{article}

\usepackage{titling} % Required for subtitle
\usepackage[english]{babel}  
\usepackage[margin=1in]{geometry}
\usepackage{setspace}
\RequirePackage[OT1]{fontenc}
\RequirePackage{graphicx}
\RequirePackage{amsthm,amssymb}
\RequirePackage[cmex10]{amsmath}
\RequirePackage[colorlinks,citecolor=blue,urlcolor=blue]{hyperref}
\numberwithin{equation}{section}

\usepackage{algorithm}
\usepackage{algpseudocode}
\usepackage{setspace}  
\usepackage[round,authoryear]{natbib}
\usepackage{graphicx,epstopdf}
\usepackage{listing}
\usepackage{amsfonts}
\usepackage{amsmath}		
\usepackage{bbm} 	
\usepackage{accents}
\usepackage{amsmath,bm}
\usepackage{mathtools}
\usepackage{booktabs}
\usepackage{xcolor}
\usepackage{tabu}\usepackage{amsmath}
\usepackage{amsfonts}
\usepackage{amssymb}
\usepackage{hyperref} % For clickable email links
\usepackage{caption} 
\usepackage{verbatim}
\usepackage{multirow}
\usepackage{authblk}
\usepackage[multiple]{footmisc}
\usepackage{easy-todo}
\usepackage{float}

\usepackage{color}
\usepackage{array}
\definecolor{dkgreen}{rgb}{0,0.6,0}
\definecolor{gray}{rgb}{0.5,0.5,0.5}
\definecolor{mauve}{rgb}{0.58,0,0.82}

\usepackage{xr}
\makeatletter
\newcommand*\rel@kern[1]{\kern#1\dimexpr\macc@kerna}
\newcommand*\widebar[1]{%
  \begingroup
  \def\mathaccent##1##2{%
    \rel@kern{0.8}%
    \overline{\rel@kern{-0.8}\macc@nucleus\rel@kern{0.2}}%
    \rel@kern{-0.2}%
  }%
  \macc@depth\@ne
  \let\math@bgroup\@empty \let\math@egroup\macc@set@skewchar
  \mathsurround\z@ \frozen@everymath{\mathgroup\macc@group\relax}%
  \macc@set@skewchar\relax
  \let\mathaccentV\macc@nested@a
  \macc@nested@a\relax111{#1}%
  \endgroup
}
\makeatother

\newtheorem{assumption}{Assumption}

\newtheorem{proposition}{Proposition}

\newtheorem{theorem}{Theorem}

\newtheorem{remark}{Remark}[section]
\newcommand{\defeq}{\equiv}
\newcommand{\diag}{\textrm{diag}}
\newcommand{\trans}{\top}              % transpose
\newcommand{\indicator}{\mathbbm{1}}     % indicator
\newcommand{\dd}{\mathrm{d}}             % derivative
\newcommand{\mfzero}{\mathbf{0}}         % zero vector
\newcommand{\mfone}{\mathbf{1}}
\newcommand{\SR}{\mathbb{R}}             % real value space
\newcommand{\SN}{\mathbb{N}}             % natural number space
\newcommand{\SY}{\mathcal{Y}}
\newcommand{\SQ}{\mathcal{Q}}

\newcommand{\wto}{\Rightarrow}

\newcommand{\op}{o_{\mathrm{P}}}         
\newcommand{\filtr}{\mathcal{F}}

\newcommand{\Exp}{\mathbb{E}}
\newcommand{\policy}{\psi}
\newcommand{\Spo}{\Psi}
\newcommand{\policyb}{\boldsymbol{\policy}}
\newcommand{\outfun}{\Upsilon}
\DeclareMathOperator*{\argmax}{arg\,max}

\newcommand{\f}{f}
\newcommand{\g}{g}
\newcommand{\p}{p}
\newcommand{\h}{h}
\newcommand{\q}{q}
\newcommand{\vq}{\boldsymbol{\q}}
\newcommand{\elbo}{J}

\newcommand{\prob}{\mathrm{Prob}}
\newcommand{\lr}{\Lambda}

\newcommand{\pint}{\theta}
\newcommand{\pintb}{\boldsymbol{\pint}}

\newcommand{\lpint}{h}
\newcommand{\lpintb}{\boldsymbol{\lpint}}
\newcommand{\lpintab}{\lpint_{a b}}
\newcommand{\pz}{\pi}
\newcommand{\pzb}{\boldsymbol{\pz}}
\newcommand{\logiz}{\omega}
\newcommand{\logizb}{\boldsymbol{\logiz}}
\newcommand{\llogiz}{\nu}
\newcommand{\llogizb}{\boldsymbol{\llogiz}}
\newcommand{\n}{n}
\newcommand{\m}{m}
\newcommand{\mn}{\m}
\newcommand{\vm}{\boldsymbol{\m}}

\newcommand{\Y}{Y}
\newcommand{\Z}{Z}
\newcommand{\vY}{\boldsymbol{\Y}}
\newcommand{\vZ}{\boldsymbol{\Z}}

\newcommand{\vI}{\boldsymbol{I}}
\newcommand{\Zjl}{Z_{i_1(j)}}
\newcommand{\Zjr}{Z_{i_2(j)}}
\newcommand{\qjl}{q_{i_1(j)}}
\newcommand{\qjr}{q_{i_2(j)}}
\newcommand{\score}{\dot{\ell}}
\newcommand{\CS}{\Delta}
\newcommand{\CSb}{\boldsymbol{\CS}}

\newcommand{\QV}{\mathcal{Q}}
\newcommand{\QVb}{\boldsymbol{\QV}}
\newcommand{\FI}{\mathcal{I}}
\newcommand{\rate}{\lambda}
\newcommand{\se}{se}
\newcommand{\bound}{d}
\newcommand{\e}{e}
\newcommand{\ve}{\boldsymbol{\e}}
\newcommand{\vR}{\boldsymbol{R}}
\newcommand{\modula}{Q}
\newcommand{\modulafun}{F}
\newcommand{\vO}{\boldsymbol{O}}
\newcommand{\vX}{\boldsymbol{X}}
\newcommand{\vmoment}{S}
\newcommand{\vmomentb}{\boldsymbol{\vmoment}}

\newcommand{\vA}{\boldsymbol{A}}
\newcommand{\vB}{\boldsymbol{B}}
\newcommand{\MLE}{\mathrm{MLE}}
\newcommand{\VE}{\mathrm{VE}}
\newcommand{\x}{x}
\newcommand{\post}{\tilde}
\newcommand{\turnover}{\xi}
\graphicspath {{figures/}}

\makeatother

\title{Batched Adaptive Network Formation} 

\author{
    Yan Xu\thanks{Assistant Professor, Department of Marketing, Virginia Tech. e-mail: 
    \protect\href{mailto:yanx19@vt.edu}{yanx19@vt.edu}.}  
    ~and  
    Bo Zhou\thanks{Assistant Professor, Department of Economics, Virginia Tech. e-mail: 
    \protect\href{mailto:bzhou@vt.edu}{bzhou@vt.edu}.}
}

\date{July 2025}

\begin{document}
\maketitle

\vspace{-0.2in}
\begin{abstract}
\noindent 
Networks are central to many economic and organizational applications, including workplace team formation, social platform recommendations, and classroom friendship development. In these settings, networks are modeled as graphs, with agents as nodes, agent pairs as edges, and edge weights capturing pairwise production or interaction outcomes. 
This paper develops an adaptive, or \textit{online}, policy that learns to form increasingly effective networks as data accumulates over time, progressively improving total network output measured by the sum of edge weights.

Our approach builds on the weighted stochastic block model (WSBM), which captures agents' unobservable heterogeneity through discrete latent types and models their complementarities in a flexible, nonparametric manner. We frame the online network formation problem as a non-standard \textit{batched multi-armed bandit}, where each type pair corresponds to an arm, and pairwise reward depends on type complementarity. This strikes a balance between exploration---learning latent types and complementarities---and exploitation---forming high-weighted networks. We establish two key results: a \textit{batched local asymptotic normality} result for the WSBM and an asymptotic equivalence between maximum likelihood and variational estimates of the intractable likelihood. Together, they provide a theoretical foundation for treating variational estimates as normal signals, enabling principled Bayesian updating across batches. The resulting posteriors are then incorporated into a tailored maximum-weight matching problem to determine the policy for the next batch. Simulations show that our algorithm substantially improves outcomes within a few batches, yields increasingly accurate parameter estimates, and remains effective even in nonstationary settings with evolving agent pools.

%\textbf{JEL classification:} 
% C12, C14

\vspace{0.2in} % Adjust spacing as needed
\end{abstract}

\section{Introduction} \label{sec:introduction}
Networks play a central role in many economic and management applications and are often modeled as \textit{graphs}, with agents as \textit{nodes}, their connections as \textit{edges}, and pairwise production or interaction outcomes as \textit{edge weights}. For example, a manager forms teams of workers to complete tasks, with team output reflected in edge weights. Similarly, a teacher seats students in pairs, or a social platform recommends connections, both aiming to foster friendships or social ties, where the success or failure of these efforts is recorded as binary edge weights. In all these cases, the decision maker---be it a manager, platform, or teacher---faces the same core challenge: how to (re)allocate effective connections that maximize total network production, measured by the sum of edge weights. 

The effectiveness of these connections hinges on complementarities between agents, that is, which workers collaborate productively or which students or users are likely to form successful friendships or social ties.\footnote{We use the term \textit{complementarity} to refer to the interaction effects between heterogeneous agents that drive pairwise outcomes. These effects may reflect productivity complementarities in teams or affinity in social tie formation. We adopt a unified terminology across applications for clarity.} In practice, the decision maker must first learn from data: by forming connections, observing outcomes, and inferring agents' unobserved heterogeneity and complementarities---the \textit{exploration} phase. Once sufficient information is gathered, a network formation policy can be determined to reallocate connections for improved overall network production---the \textit{exploitation} phase. \textit{Offline} policies that carry out the two phases in a single round often perform poorly: updating the policy too early with insufficient data risks inaccurate estimates, while delaying too long may miss valuable opportunities for improvement. Both scenarios reflect an imbalance in the exploration–exploitation trade-off, resulting in suboptimal outcomes. 

A more effective approach is to adaptively update the policy as new data become available, dynamically balancing exploration and exploitation. In this paper, we propose such an adaptive, or \textit{online}, policy designed to form increasingly productive networks. We frame this as a \textit{Batched Multi-Armed Bandit (Batched MAB)} problem, where each possible pair of agent latent types corresponds to an arm. The algorithm proceeds in batches: after each batch of observed outcomes, it updates beliefs about complementarities and uses these to solve a maximum-weight matching problem to determine the next batch’s network. This updated policy is applied to pair agents for improved outcomes, with the resulting data fed into the next round of inference and optimization. This iterative structure enables the algorithm to progressively refine its decisions, yielding continuous improvement over time.

One practical challenge decision makers often face is that only edge-level outcomes (e.g., the success of teamwork, the development of friendships or social ties) are observable, while individual contributions typically remain unobserved. Moreover, these outcomes depend on the agents' unobserved characteristics (e.g., workers' soft skills,  students' or platform users' personalities) often in complex, nonlinear patterns not amenable to parametric specification.\footnote{The importance of unobserved agent heterogeneity on complementarities has been documented across various contexts, including teamwork (\citet{weidmann2021team}, \citet{weigel2024supermodular}, and \citet{xu2024heterogeneous}), medical matching (\citet{agarwal2015empirical}), and online dating (\citet{lee2016effect}).}  These challenges motivate our adoption of a weighted stochastic block model (WSBM), which captures agents' unobserved heterogeneity through $K$ discrete latent types (see, e.g., \citet{bonhomme2021teams} and \cite{jochmans2024nonparametric}). The expected edge-level outcome---reflecting complementarities between agents---depends on the latent types of the agents in each agent pair and is parameterized by $\pintb\in\SR^{K\times K}$. A key advantage of this WSBM-based framework is that it imposes no functional form restrictions on $\pintb$, which allows for more flexible modeling of heterogeneous complementarities compared to production models that that impose structure, such as strongest/weakest-link specifications (see, e.g., \citet{JohKamKriLi2018Exploration}), linear models (\cite{jochmans2019fixed}), or nonlinear models with specific functional forms (e.g., the constant elasticity of substitution form adopted by \citet{ahmadpoor2019decoding}).\footnote{Both \citet{jochmans2019fixed} and \citet{ahmadpoor2019decoding} model agents’ unobserved heterogeneity as continuous-valued fixed effects, whereas the WSBM framework captures it through $K$ discrete latent types. Due to this difference, the comparison between our framework and these two works is not entirely direct.}

To rigorously integrate WSBM-based network models into a tractable online decision-making framework, we build on H\'ajek-Le Cam's theory of convergence of statistical experiments, drawing in particular on the recent extension by \cite{hirano2023asymptotic} for batched adaptive experiments. We establish a \textit{batched local asymptotic normality (batched LAN)} result for the WSBM-based network production model, assuming that the network formation policies are determined prior to each batch. This result shows that, within each batch, the complex network production experiment can be asymptotically represented by a much simpler \textit{limit experiment} involving a single draw from a normal distribution. Subsequently, the entire $T$-batch network experiment can be approximated by a $T$-stage Gaussian bandit environment, each stage generates an independent Gaussian signal, with mean determined by $\pintb$ and variance depending on the network formation policy employed in the corresponding batch.\footnote{In particular, type pairs that are sampled more frequently within a batch yield more precise parameter estimates (i.e., signals with lower variance); see Section~\ref{sec:asmptotic_results} for details.} Under Gaussianity, these per-batch signals serve as maximum likelihood estimates (MLEs) in the limit.

Nevertheless, computing the finite-sample MLE is generally infeasible for WSBM models due to the intractability of their likelihoods (see \citet[Chapter 5]{graham2020econometric} and \citet{bonhomme2021teams}). To address this, we adopt a variational estimation approach, applied independently to each batch, and show that it is asymptotically equivalent to the MLE. In addition to estimating $\pintb$, the variational method also produces local variational approximations (LVAs) as a valuable by-product for inferring agents' latent types. These LVAs, given as categorical distributions over the $K$ possible types, approximate the posterior distribution of the latent type variable $\vZ$, and are shown via simulation to rapidly concentrate around true latent types. All these results are developed under \textit{batched sparse network asymptotics}, reflecting practical constraints---for example, that an agent may participate in only a limited number of tasks per batch (e.g., one month), regardless of how many potential collaborators they have.\footnote{Specifically, the batched sparse asymptotics allows the expected \textit{degree} (i.e., the number of edges per node) in each batch to grow as slowly as logarithmically in the number of nodes.}

These theoretical results convey a simple yet powerful message: batched variational estimates can be treated as Gaussian signals for the production parameter $\pintb$, while the batched LVAs can be viewed as categorical signals for agents' latent types $\vZ$. Building on this insight, we aggregate information on $\pintb$ and $\vZ$ across batches using standard Bayesian methods. From the resulting posteriors, we either compute point estimates such as posterior means in a \textit{greedy} manner, or sample parameter values following the \textit{Thompson sampling} principle. These estimates or sampled values are then incorporated into a constrained \textit{maximum-weight matching} problem, the solution of which determines the network formation policy for the next batch. In particular, we propose a \textit{Hybrid Greedy-Thompson (HGT)} algorithm, which uses posterior mean as the point estimate for the production parameter (``greedy-in-$\pintb$'') and samples agent-type realizations from the LVAs (``Thompson-in-$\vZ$''). Simulation results show that our HGT algorithm substantially improves overall outcomes within a few batches, while simultaneously refining estimates for both $\pintb$ and $\vZ$ over time. We further evaluate performance in non-stationary environments with agent entry and exit. In these settings, HGT algorithm remains robust, though requiring a few more batches to achieve comparable performance.

% Building on this insight, we design our online network formation policy. Specifically, we treat each batch’s variational estimate as a noisy signal and apply Bayesian updating to form a posterior distribution for $\pintb$.
%Specifically, we apply Bayesian updating to obtain a posterior distribution for $\pintb$ using all completed batches. 
% Similarly, we update the batched LVAs in the form of categorical distributions to derive the posterior distributions of agents' latent types. We then compute a point estimate for $\pintb$ by taking the posterior mean and generating realizations of agents' latent types by sampling from the LVA posteriors. With these estimates and realizations, we then formulate and solve a (possibly constrained) matching problem to maximize the edge weights (or outcomes), which determines the network formation policy for the next batch. Simulation results show that our algorithm can substantially improve overall outcomes within just a few batches. Moreover, our adaptive algorithm yields increasingly accurate estimates for both the production parameter $\pintb$ and the agents' latent types over time and achieves significant gains (i.e., lower regret). 
%Additionally, our approach achieves significant gains (i.e., lower regret) compared to random network formation, heuristic-based assignment, and assignments based on the previously discussed offline policy. 
% Finally, we show that the algorithm remains robust in non-stationary environments, where agents may enter and exit the network.

\subsubsection*{Related literature}

Our paper relates to the extensive literature on network models; for a comprehensive review, see \citet{graham2020network}. In particular, we contribute to the growing body of work on network production models based on the weighted stochastic block model (WSBM) (see, e.g., \cite{bonhomme2021teams}, \cite{jochmans2024nonparametric}), which flexibly capture ``nonparametric'' complementarities among heterogeneous agents. These models have been applied in recent empirical studies such as \citet{weigel2024supermodular} (tax collector teams) and \citet{xu2024heterogeneous} (sales-force teams) to estimate heterogeneous complementarities and evaluate offline network formation policies. We extend this literature by developing an online framework in which network formation policies are updated adaptively, enabling continuous improvement in production outcomes over time.

While motivated by the online setting, our theoretical results---namely, the batched local asymptotic normality (LAN) for WSBM-based network production models and the asymptotic equivalence of variational estimators to maximum likelihood estimators under sparse network asymptotics---also readily apply to offline estimation by treating the full sample as a single batch. This work extends the asymptotic theory of feasible variational inference (consistency and asymptotic normality), established for classical (unweighted) stochastic block models (SBMs; see, e.g., \cite{bickel2009nonparametric}, \cite{bickel2013asymptotic}), to weighted, exogenously formed network SBMs.

We also contribute to the literature on multi-armed bandit (MAB) algorithms (e.g., \cite{thompson1933likelihood}, \cite{lai1985asymptotically}, \cite{auer2002using}, \cite{kasy2021adaptive}; and \cite{cesa2023adaptive} for applications in economics), particularly in the context of team formation and assignment problems (\cite{johari2018exploration}, \cite{johari2021matching}, \cite{eichhorn2022online}). Incorporating WSBM into this setting enables flexible modeling of complementarities between heterogeneous agents whose characteristics are unobserved. However, it introduces two key departures from standard MAB formulations. First, the arms---here, type pairs---are not directly observed, but instead their inferred probabilities through local variational approximations. Second, the network formation policy are solutions to constrained optimization problems, rather than explicit functions of simple descriptive statistics. The first departure is specific to our WSBM-based setting, while the second connects our work to the \textit{combinatorial semi-bandit} literature, in which feasible policies often involve selecting matchings subject to combinatorial constraints (e.g., \cite{audibert2014regret}, \cite{chen2013combinatorial}, and \cite{kasy2023matching}). To address these challenges, we adopt the batched bandit framework (e.g., \cite{perchet2016batched}, \cite{zhang2020inference}) and leverage Le Cam’s asymptotic theory for batched adaptive experiments by \cite{hirano2023asymptotic}, enabling the integration of complex network econometric modeling into online decision-making.

The rest of the paper is organized as follows. Section~\ref{sec:onlinenetworkformation} introduces the WSBM-based network model and frames the online network formation problem as a batched MAB. Section~\ref{sec:estimation} presents our variational approximation estimation strategy, followed by asymptotic analysis in Section~\ref{sec:asmptotic_results}. Section~\ref{sec:onlinnetworkformation_detail} details our algorithm, including Bayesian updating and an agent-level constrained maximum-weight matching problem, making it readily applicable to practical settings. Section~\ref{sec:simulation} presents a comprehensive Monte Carlo study demonstrating the effectiveness of our algorithm. Section~\ref{sec:conclusion} concludes.

% Complementarity & matching \cite{graham2020econometric}

\section{Online Network Formation} \label{sec:onlinenetworkformation}

\subsection{Setup} \label{subsec:setup}
Consider a pool of agents indexed by $1, \dots, \n$, represented as \textit{nodes} in a graph. Each agent belongs to one of $K$ latent types, which capture heterogeneity in agents' unobserved characteristics. The latent type of agent $i$ is recorded in the random variable $Z_i$, which takes values in $\{1,\dots,K\}$. We assume $Z_i$ are independently and identically distributed (i.i.d.) with probabilities
\begin{align}
\prob(Z_i = a) = \pz(a),
\end{align}
for $a = 1,\dots,K$, where $\pz(1) + \cdots + \pz(K) = 1$. 

Agents are assigned into $m$ pairs, represented as \textit{edges} in the graph, to complete $m$ tasks.\footnote{We use the term ``task'' broadly to refer to the intended outcome of a pairwise assignment, including both collaborative activities and social interactions---e.g., teaming agents for joint work, seating students to foster friendships, or recommending platform users to follow each other.} Each pair $j \in \{1, \dots, \m\}$ is formed by two distinct agents $i_1(j), i_2(j) \in \{1, \dots, \n\}$ with $i_1(j) \neq i_2(j)$. Throughout, we focus exclusively on two-agent collaborations and therefore use ``network formation'' and ``agent pairing'' interchangeably.\footnote{Extending our framework to collaborations involving more than two agents would require hypergraph theory and generalization of the asymptotic results in Section~\ref{sec:estimation}, which we leave for future work.}

The outcome of the Task $j$, completed by Pair $j$, is modeled as the \textit{weight} of the corresponding edge in the graph and is captured by a random variable $Y_j$. The weight $Y_j$ can be binary, where $Y_j = 1$ indicates success and $Y_j = 0$ indicates failure, or continuous, where $Y_j \in \SR$. We assume that the outcomes $Y_j$ are i.i.d.\ with distribution 
\begin{align} \label{eqn:model_pairwise}
\p_{\pint}\left(Y_j \,|\, \Zjl, \Zjr\right),
\end{align}
which depends on the type pair, $\left(\Zjl,\Zjr\right)$, where $\pint = \pint_{(\Zjl,\Zjr)} \in\SR$ is an unknown parameter. For simplicity, in this pairwise production model $\p_{\pint}(\cdot|a,b)$, we restrict our analysis to the univariate case, where $\pint_{a b}$ is a scalar for each type pair $(a,b)$, with $a,b = 1,\dots,K$ (abbreviating $\pint_{a b}$ as $\pint$ when used as a subscript). However, our results extend naturally to the multivariate, albeit with more involved notation. We impose no functional form restrictions on $\p_{\pint}(\cdot|a,b)$ beyond mild regularity conditions (required for Assumption~\ref{assm:LAN_individual} to hold).

\begin{remark}
Covariates, when available, can be incorporated as additional inputs in the pairwise production model by specifying an appropriate functional form. That is, the location model in (\ref{eqn:model_pairwise}) can be extended to, for example, a linear regression for continuous $Y_j$ or a logistic regression for binary $Y_j$; see \cite{bonhomme2021teams} and \cite{xu2024heterogeneous}. The asymptotic results in Section~\ref{sec:asmptotic_results} are conjectured to continue to hold, provided the extended pairwise model satisfies Assumption~\ref{assm:LAN_individual} and the first-order terms---the so-called central sequences---associated with $\pint_{a b}$ are asymptotically independent of those associated with the covariate parameters (as is the case for linear or logistic regression). This condition ensures that our asymptotic analysis can proceed under the ``as if'' assumption that the covariate parameters are known.  
\end{remark}

\begin{remark}
In this paper, we do not consider different types of tasks (i.e., task heterogeneity) and thus do not explicitly model the matching problem between agent types and task types. For studies addressing this aspect, see, e.g., \citet{johari2021matching}.
\end{remark}

\begin{remark}
Although we do not impose it in this model setup, one could additionally assume the symmetry condition $\p_{\pint}(\cdot \,|\, a,b) = \p_{\pint}(\cdot \,|\, b,a)$ for all $a,b=1,\dots,K$, implying that the order of agents in a pair is interchangeable and does not affect outcomes. This symmetry condition corresponds to settings in which the roles of the two agents are functionally indistinguishable. For instance, in teamwork applications, it may not matter which agent is listed first; and in the context of social tie formation, when mutual connections are the outcome of interest, user ordering is irrelevant.
\end{remark}

\subsection{Network Formation as an Optimization Problem}
To motivate the policy design problem, we begin with an oracle setting in which the agents’ latent types and the pairwise production model are known to the decision maker. The decision maker's objective is to form agent pairs that maximize total network output. Specifically, the decision maker's objective is to form agent pairs that maximize total network output. Specifically, she/he aims to solve
\begin{align} \label{eqn:optimization_oracle}
\max_{\policy\in\Spo} \, \policy(a,b) \, \outfun(a,b)
\end{align}
where $\outfun(a,b) \defeq \Exp_{\pint}\left(Y_j \,|\, \Zjl = a, \Zjr = b\right)$ is the expected production of a type $(a, b)$ pair, and the \textit{pairing policy} function $\policy(a,b)$ determines the relative frequency (or probability mass) with which each type pair $(a, b)$ is assigned across $\m$ pairs. 
 
The unrestricted policy space $\Spo$, to which $\policy$ belongs, comprises all valid probability distributions over $K\times K$ possible type pairs:
\begin{align}
\Spo = \left\{A \in [0, 1]^{K\times K} \,\bigg|\, \sum_{a=1}^{K}\sum_{b=1}^{K}A[a,b] = 1\right\}.
\end{align}
That is, each $\policy(a,b)$ as type-$(a,b)$ pair probability lies in $[0, 1]$, and all probabilities sum to one. In real-world scenarios, the optimization problem often incorporates additional constraints. For instance, in the teamwork example, a key restriction is the workload constraint (see \cite{weigel2024supermodular} and \cite{xu2024heterogeneous}) which ensures a fair distribution of tasks across workers. Another commonly imposed constraint in bandit literature is the ``clipping constraint,'' designed to avoid sampling probabilities that are too small and that would lead to inconsistent parameter estimates. We formally state these two constraints as follows.
\begin{itemize}
\item[-] \textit{The workload constraint:} Each agent can participate in only a limited number of tasks, specifically between $[\bound_l, \bound_h]$, where $\bound_l \leq \bound_h \in \SN^{+}$, within a given period. This agent-level constraint induces the policy-level constraint: for a total of $\m$ pairs,
\begin{align} \label{eqn:constraint_workload}
\sum_{b=1}^{K}(\policy(a,b) + \policy(b,a)) \in \frac{\n_a}{\m}\times[\bound_l, \bound_h], \textrm{~~for~all~} a=1,\dots,K, 
\end{align}
where $\n_a = \sum_{i=1}^{\n}\indicator\{\Z_i = a\}$ counts the total number of type-$a$ agents. 
\item[-] \textit{The clipping constraint:} For each type pair $(a, b)$, where $a,b=1,\dots,K$, the sampling probability satisfies
\begin{align} \label{eqn:constraint_clipping} 
\policy(a,b) \in [\rate, 1 - \rate], 
\end{align}
for some clipping rate $\rate \in (0,0.5)$. 
\end{itemize}
Consequently, the optimization problem becomes
\begin{align} \label{eqn:optimization_oracle_constrained}
\max_{\policy\in\Spo^{c}} \, \policy(a,b) \, \outfun(a,b),
\end{align}
where $\Spo^{c} = \left\{\policy \in \Spo \,\big|\, \policy \textrm{~satisfies~} (\ref{eqn:constraint_workload}) \textrm{~and~} (\ref{eqn:constraint_clipping}) \right\}$ denotes the constrained policy space. In this oracle scenario where the agent types $\vZ = (Z_1,\dots,Z_{\n})^\trans$ are observed and the pairwise production models $\p_{\pint}(\cdot | a,b)$ are known, this optimization problem can be solved readily via linear programming. 

In practice, however, neither the agents' types nor the pairwise production models are known to the decision maker. This necessitates a two-phase procedure: first, estimate these parameters using data from sampled agent pairings and their outcomes---the exploration phase; then, use the obtained estimates to solve the network output maximization problem (\ref{eqn:optimization_oracle_constrained})---the exploitation phase. When this process is carried out in a single round, it is known as an \textit{offline} policy, where the decision maker commits to a fixed policy without collecting new data to refine the policy further. A more effective strategy is to run adaptive experiments, in which the agent-pairing policy is updated sequentially using newly collected data. This iterative process, referred to as an \textit{online policy}, balances exploration (learning parameters) and exploitation (maximizing network output based on current beliefs). In the following subsection, we cast this iterative learn-and-earn procedure as a non-standard Batched Multi-Armed Bandit problem and develop an algorithm specifically tailored for WSBM-based adaptive network formation.

\subsection{Online Network Formation as Batched Bandits}
In the classical Multi-Armed Bandit (MAB) framework, a decision maker sequentially selects from a set of treatments---akin to a gambler choosing among slot machine arms---each associated with an unknown reward distribution. The objective is to maximize cumulative rewards over time by balancing exploration (sampling different arms to learn their reward distributions) and exploitation (favoring the arm currently believed to yield the highest expected return). In our setting, each agent-type pair $(a, b)$ represents an arm, and edge-level outcomes serve as stochastic rewards governed by the conditional distribution $\p_{\pint}(\cdot|a, b)$. However, the WSBM-based network formation problem is inherently more complex, departing from the standard MAB paradigm in two fundamental ways.

First, in addition to the production model $\p_{\pint}$, the agent types $\vZ$ are themselves unobserved. This introduces a key departure from the standard setting: when assigning a pair, the underlying type combination---and thus the specific arm being pulled---is unknown. Instead, decisions must rely on the posterior distribution of $\vZ$, obtained via the variational approximation (see Section~\ref{sec:estimation}), which induces corresponding probabilities over type combinations. As data accumulate, this posterior distribution becomes increasingly concentrated around the true latent types, thereby improving the decision-maker’s understanding of each agent and enabling better-informed pairing decisions.

Second, the development of the sampling policy that governs arm-pulling probabilities differs significantly from that of a standard MAB problem, where policies are often explicit functions of descriptive statistics such as accumulated rewards and arm-pulling frequencies. These functions are typically guided by inferences such as the posterior distribution, point estimates, or upper confidence bounds of the unknown parameters, corresponding to Thompson sampling (\cite{thompson1933likelihood}), greedy heuristics, or UCB algorithms (\cite{lai1985asymptotically, auer2002using}), respectively. In contrast, our network formation policy is determined by solving the optimization problem in (\ref{eqn:optimization_oracle}), where the unknown parameters are replaced with their inferred values. This procedure is further complicated by practical restrictions such as the workload constraint.

To tackle these challenges, we formulate our algorithm in a \textit{batched} bandit setting, where the network formation policy $\policy$ is updated at discrete time intervals. Specifically, we carry out the network formation and production process over $T$ batches for a potentially evolving group of agents or their types. 
In each batch $t \in \{1,\dots,T\}$, we assign $\n_t$ agents into $\m_t$ pairs according to a policy $\policy_t(a,b)$; we record the pair assignments in mappings $(i_{t,1}(j), i_{t,2}(j))$ and the corresponding production outcomes in variable $Y_{t,j}$, for $j = 1,\dots,\m_t$. Both $T$ and $\m_t$ are either predetermined or exogenous, and thus treated as deterministic.\footnote{In real-world applications, both the batched decision making and the number of tasks in each batch, $\m_t$, can naturally arise from institutional or market constraints. For instance, the real estate team setting described in \cite{xu2024heterogeneous} is well suited to a batched implementation, where the manager could periodically update the assignment policy (e.g., monthly or quarterly), and the number of properties $\m_t$ to be managed in each period is exogenously determined by market availability. Similar batching structures apply in other domains such as education \cite{rohrer2021proximity} and online platforms \cite{rajkumar2022causal}.} Importantly, we require that each policy $\policy_t$ be specified prior to batch $t$, based solely on information available from the previous batches. We formalize this ``pre-determined policy'' condition in the following assumption.

\begin{assumption}[Pre-determined Policy] \label{assm:adaptive_policy}
For each batch $t = 1,\dots,T$, let variables $\vI_t = ((i_{t,1}(1),i_{t,2}(1))^\trans,\dots,(i_{t,1}(m_t),i_{t,2}(m_t))^\trans)^\trans$ and $\vY_t = (Y_{t,1},\dots,Y_{t,\m_t})^\trans$ collect the network formation and production outcomes, respectively. Define the accumulated information available prior to batch $t$ as $\sigma$-algebra $$\filtr_{t-1} \defeq \sigma\left(\vI_{s}, \vY_{s}, s = 1,\dots,t-1\right).$$ We assume that the policy $\policy_t \in \Spo^{c}$ is measurable with respect to $\filtr_{t-1}$. 
\end{assumption} 

The pre-determined policy condition in Assumption~\ref{assm:adaptive_policy} is naturally satisfied by any feasible network formation algorithm that determines the pair assignment policy for each batch based solely on data from previous batches, and keeps that policy fixed throughout the batch. This condition allows us to treat the network formation $\vI_t$ for batch $t$ as fixed, or nonrandom. In settings where agents form pairs themselves, this condition implies the Network Exogeneity assumption (see \citet[Assumption~1]{bonhomme2021teams}), which requires that network formation be independent of agent-pair-specific shocks, thereby ruling out cases where agents have prior knowledge of such shocks and use it to influence pair formation.\footnote{Relaxing this condition in such cases requires explicitly modeling the network formation process; see also \cite{bonhomme2021teams}.} In our setting, however, this concern does not arise, as all pairings are assigned by a decision maker according to pre-specified policies. 

We conclude this section with a brief overview of our online network formation policy. The complete procedure is detailed in Algorithm~\ref{algorithm:HGT} in Section~\ref{sec:onlinnetworkformation_detail}. Its core components---including  variational estimation, asymptotic results, Bayesian updating, and an agent-level optimization problem---are presented in Sections~\ref{sec:estimation}--\ref{sec:onlinnetworkformation_detail}.

\bigskip
\noindent
\textbf{Online Network Formation Policy:}
Starting with an initial policy $\policy_1$, the algorithm proceeds iteratively over batches $t = 1, \dots, T$ according to the following procedure:
\begin{itemize}
\item[1)] \textit{Network Formation and Production:} Agents are matched into pairs according to the current policy $\policy_t$. The resulting pair assignments and production outcomes are recorded as $(\vI_t,\vY_t)$.
\item[2)] \textit{Estimation:} Given the observed data $(\vI_t, \vY_t)$, we estimate production parameters $\hat{\pintb}_t$ and the latent agent characteristics $\hat{\vZ}_t$ via a variational approach.
\item[3)] \textit{Updating:} Treating $\hat\pintb_{t}$ and $\hat\vZ_{t}$ as independent signals from batch $t$ about $\pintb$ and $\vZ$, respectively, we apply Bayesian updating to obtain their posterior distributions. 
\item[4)] \textit{Optimization:} From the updated posteriors, we obtain point estimates or posterior draws for $\pintb$ and $\vZ$, which are then used to solve the constrained optimization problem in equation~\eqref{eqn:optimization_oracle_constrained}, yielding the policy $\policy_{t+1}$ for the next batch.
\end{itemize}

% \begin{algorithm}[H]\label{algorithm_onlineteamformation}
% \caption{Online Network Formation Policy} \label{algorithm_onlineteamformation}

% \textbf{Require:} Number of batches $T$, number of agents $n$, initial pairing constraints, and prior belief over types and production parameters.

% \vspace{0.5em}
% \quad \textit{Initialize:} Assign pairs for batch $t=1$ randomly, subject to constraints.

% \vspace{0.5em}
% \textit{For} $t = 2, \dots, T$:

% \quad \textit{Estimation:} Use data $\mathcal{F}_{t-1}$ to obtain estimates of latent types $\hat{\vZ}_{t-1}$ and production parameters $\hat{\theta}_{t-1}$ via Expectation-Maximization Algorithm.

% \quad \textit{Updating:} Apply Bayesian updates as described in Section~\ref{sec:onlinnetworkformation_detail}, treating each batch's estimates as independent signals. Denote updated posteriors as $\hat{\theta}_{1:t-1}$ and $\hat{\vZ}_{1:t-1}$.

% \quad \textit{Optimization:} Solve the assignment problem in (\ref{eqn:optimization_oracle_constrained}) using the updated estimates $\hat{\theta}_{1:t-1}$ and $\hat{\vZ}_{1:t-1}$ to obtain the pairing policy $\psi_t$.
% \end{algorithm}

\section{Estimation via Variational Approximation} \label{sec:estimation}
In this section, we introduce our estimation strategy based on a variational approximation, applied independently to each batch. For notational simplicity, we suppress the batch index $t$ throughout the section (e.g., for the number of agents $\n_t$, number of tasks $\m_t$, policy function $\policy_{t}$, and outcome variables $Y_{t,j}$ and $\vY_t$), unless needed for clarity.

Under the i.i.d.\ assumption for $Y_j$, the conditional likelihood of $\vY = (Y_1,\dots,Y_\m)^\trans$ given agents' latent types $\vZ$ is 
\begin{align}
\p_{\pintb}(\vY|\vZ) = \prod_{j=1}^{\m} \p_{\pint}(Y_j|\Zjl,\Zjr),
\end{align}
where the vector-valued parameter $\pintb\in\SR^{K\times K}$ collects all pair-specific production parameters $\pint = \pint_{a b}$ (of pairwise model $\p_{\pint}(\cdot|a,b)$), $a,b=1,\dots,K$. The types of the agents assigned to task $j$, $(\Zjl,\Zjr)$, is from the network formation data $\vI$. While $\vI$ determines the structure of the likelihood, it is assumed to be pre-determined and treated as fixed (i.e., non-random), and is thus omitted from the notation. The likelihood of $(\vY,\vZ)$ is then given by
\begin{align}
\f_{\pzb,\pintb}(\vY,\vZ) = \p_{\pzb}(\vZ)\p_{\pintb}(\vY|\vZ),
\end{align}
where $\p_{\pzb}(\vZ) = \prod_{i=1}^{\n}\pz(Z_i)$ denotes the marginal likelihood of $\vZ$. Following \citet{bickel2013asymptotic}, we refer to the model in which $\vZ$ is observed as the \textit{complete graph model}.

In the complete graph model, the type probabilities $\pzb$ and production parameters $\pintb$ can be directly estimated by maximizing the joint likelihood $\f_{\pzb,\pintb}(\vY,\vZ)$. However, when agent types $\vZ$ are unobserved, maximum likelihood estimation becomes computationally intractable, as it requires working with the marginal likelihood $\g_{\pzb,\pintb}(\vY)$, computed by summing over all possible type assignment vectors $\ve = (\e_1,\dots,\e_\n)^\trans$, where each $\e_i\in\{1,\dots,K\}$:
\begin{align}
\g_{\pzb,\pintb}(\vY) = \sum_{\e_1=1}^{K}\cdots\sum_{\e_\n=1}^{K}\f_{\pzb,\pintb}(\vY,\ve).
\end{align}
This summation takes $K^{\n}$ operations and quickly becomes computationally infeasible as $n$ grows. To circumvent this challenge, we employ a \textit{mean-field variational approximation} to the marginal likelihood below, a technique widely used in latent variable models to enable tractable inference.

Write the joint likelihood as $\f(\vY,\vZ) = \h(\vZ|\vY)\g(\vY)$, we approximate the conditional likelihood $\h(\vZ|\vY)$ with a factorized term, $\q(\vZ) \defeq \prod_{i}q_i(Z_i)$. Each component $q_i$, referred to as a \textit{local variational approximation (LVA)}, serves as an individual approximation to the posterior distribution for the latent variable $Z_i$ over the support $\{1,\dots,K\}$. The discrepancy between $\q(\vZ)$ and $\h(\vZ|\vY)$ is measured by the Kullback–Leibler (KL) divergence, defined as
\begin{equation} \label{eqn:KL1}
D_{\mathrm{KL}}\left(\q(\vZ) \,\|\, \h(\vZ|\vY)\right) \defeq \sum_{Z_1=1}^{K}\cdots\sum_{Z_\n=1}^{K} \q(\vZ)\log\frac{\q(\vZ)}{\h(\vZ|\vY)}.
\end{equation}
Expanding this expression yields
\begin{equation} \label{eqn:KL2}
\begin{aligned}
D_{\mathrm{KL}}\left(\q(\vZ) \,\|\, \h(\vZ|\vY)\right) 
&= \sum_{Z_1=1}^{K}\cdots\sum_{Z_\n=1}^{K} \q(\vZ)\left[\log\q(\vZ) - \log\f(\vY,\vZ) + \log\g(\vY)\right] \\
&=~ \log\g(\vY) - \elbo(\vY),
\end{aligned}
\end{equation}
where
\begin{align*}
\elbo(\vY) = \elbo_{\pzb,\pintb;\vq}(\vY) \defeq \sum_{Z_1=1}^{K}\cdots\sum_{Z_\n=1}^{K}\q(\vZ)\left[\log\frac{\f_{\pzb,\pintb}(\vY,\vZ)}{\q(\vZ)}\right],
\end{align*}
and the boldface $\vq = (q_1,\dots,q_\n)$ denotes the collection of local variational approximations. The space where $\vq$ takes value is given by $$\SQ \defeq \left\{ (q_1,\dots,q_\n) \in \big([0,1]^{K}\big)^{\n} \,\big|\, q_i(1)+\cdots+q_i(K) = 1, ~ \forall i = 1,\dots,\n \right\}.$$

The term $\elbo(\vY)$ is known as the \textit{Evidence Lower Bound (ELBO)} which provides a lower bound on the log-likelihood $\log\g(\vY)$, with the observed data $\vY$ being the ``evidence''. This inequality holds because the KL-divergence is always non-negative, implying $\elbo(\vY) \leq \log\g(\vY)$. In practice, the ELBO offers a computationally tractable proxy for the generally intractable log-likelihood $\log\g(\vY)$. Specifically, it simplifies to
\begin{equation} \label{eqn:elbo}
\begin{aligned} 
\elbo_{\pzb,\pintb;\vq}(\vY) 
=&~ \sum_{i=1}^{\n}\sum_{a=1}^{K}q_{i}(a)\left(-\log\q_{i}(a) + \log\pi(a)\right) \\
&~ + \sum_{j=1}^{\m}\sum_{a=1}^{K}\sum_{b=1}^{K} \qjl(a)\qjr(b)\log \p_{\pint}(Y_j|a,b),
\end{aligned}
\end{equation}
where computing the summations now requires only $\n\times K + \m\times K^2$ operations in total. 

%On one hand, we aim to make $\elbo_{\pzb,\pintb;\vq}(\vY)$ as close as possible to the log-likelihood $\log\g_{\pzb,\pintb}(\vY)$ so that the former serves as a good approximation to the latter. This is achieved by maximizing $\elbo_{\pzb,\pintb;\vq}(\vY)$ with respect to $\vq\in\SQ$, which is treated as a newly introduced parameter. On the other hand, we also maximize $\elbo_{\pzb,\pintb;\vq}(\vY)$ with respect to the model $\pzb$ and $\pintb$ in order to obtain their estimates.  These optimization steps are naturally embedded in the framework of the \textit{Expectation-Maximization (EM)} algorithm (see, e.g., \citet{daudin2008mixture}), which proceeds as follows.

We aim to make $\elbo_{\pzb,\pintb;\vq}(\vY)$ as close as possible to the log-likelihood $\log\g_{\pzb,\pintb}(\vY)$, so that it serves as a good approximation. This is achieved by jointly maximizing $\elbo_{\pzb,\pintb;\vq}(\vY)$ with respect to $\vq$, which is treated as a newly introduced parameter. In the meantime, we optimize over the model parameters $\pzb$ and $\pintb$ to obtain their variational estimates. This joint optimization is naturally embedded in the framework of the \textit{Expectation-Maximization (EM)} algorithm (see, e.g., \citet{daudin2008mixture}), which proceeds as follows:
\begin{itemize}
\item[-] \textit{Initialization}:  Set initial values for all parameters, denoted $(\pzb^*,\pintb^*,\vq^*)$.
\item[-] \textit{Expectation Step}: Given $(\pzb^*,\pintb^*)$, update $\vq^*$ by solving
\begin{align*}
\vq^* = \argmax_{\vq\in\SQ} \, \elbo_{\pzb^*,\pintb^*;\vq}(\vY);
\end{align*}
\item[-] \textit{Maximization Step}: Given $\vq^*$, update $(\pzb^*,\pintb^*)$ by solving
\begin{align*}
(\pzb^*,\pintb^*) = \argmax_{\pzb\in\SR^{K}, \, \pintb\in\SR^{K\times K}} \, \elbo_{\pzb,\pintb;\vq^*}(\vY);
\end{align*}
\item[-] \textit{Exit}:  Iterate the Expectation and Maximization steps until convergence. 
% Denote the final outputs by $(\check\pzb,\check\pintb,\check\vq)$, referred to as the variational estimates.
\end{itemize}

\section{Asymptotic Results} \label{sec:asmptotic_results}
This section establishes the asymptotic results of the WSBM and its batched variational estimators above. We begin in Section~\ref{subsec:LAN_Z_observed} with the complete graph model, where the agent types $\vZ_t$ are observed. In this setting, we show that the complete graph model converges to a Gaussian shift experiment (in H\'ajek-Le Cam's sense), provided that each pairwise production model $\p_{\pint}\left(\cdot \,|\, a,b\right)$ satisfies a similar convergence. Section~\ref{subsec:LAN_Z_unobserved} then demonstrates that the variational estimates (based on $\elbo_{\pzb,\pintb;\vq}(\vY_t)$) are asymptotically equivalent to the complete graph MLEs (based on $\f_{\pzb,\pintb}(\vY_t,\vZ_t)$). For notational clarity, we vectorize the production parameter $\pintb$ as an element of $\SR^{K^2}$ throughout this section.

\subsection{Batched Local Asymptotic Normality} \label{subsec:LAN_Z_observed}
We first establish a \textit{batched local asymptotic normality (batched LAN)} result for the complete graph model, where the latent types $\vZ_t$ are observed. This is done under two high-level conditions on the pairwise production models $\p_{\pint}(\cdot | a,b)$ for $a,b = 1,\dots,K$, stated in the following assumption.

\begin{assumption} \label{assm:LAN_individual}
\begin{itemize}
\item[(a)] Let $Y_j$ be generated according to the individual production model $\p_{\pint}(\cdot|a,b)$. For all $\lpint\in\SR$ and $a,b=1,\dots,K$, we assume that as $\m\to\infty$,
\begin{align} \label{eqn:individual_LAN}
\log\prod_{j=1}^{\m} \frac{\p_{\pint+\lpint/\sqrt{\m}}(Y_j|a,b)}{\p_{\pint}(Y_j|a,b)} = \frac{\lpint}{\sqrt{\m}}\sum_{j=1}^{\m}\score_{\pint}(Y_j|a,b) - \frac{\lpint^2}{2}\FI_{\pint}(a,b) + \op(1),
\end{align}
where the score function $\score_{\pint}(\cdot|a,b)$ satisfies $\Exp_{\pint}\big[\score_{\pint}(Y_j|a,b)\big] = 0$, $\frac{1}{\sqrt{\m}}\sum_{j=1}^{\m}\score_{\pint}(Y_j|a,b) \wto \mathcal{N}(0,\FI_{\pint}(a,b))$, and the Fisher information $\FI_{\pint}(a,b) \defeq \Exp_{\pint}\big[\score_{\pint}^2(Y_j|a,b)\big]$ is finite.
\item[(b)] For all $a,b=1,\dots,K$, denote by $\hat\pint_{a b}^\MLE$ the maximum likelihood estimator (MLE) for $\pint_{a b}$, we assume $\sqrt{\m}(\hat\pint_{a b}^\MLE - \pint_{a b}) \wto \mathcal{N}(0, \FI_{\pint}^{-1}(a,b))$.
\end{itemize}
\end{assumption}

Assumption~\ref{assm:LAN_individual}(a) assumes that the production model for each type pair $(a,b)$ is locally asymptotically normal (LAN); see, e.g., \citet[Definition~7.14]{van2000asymptotic}. That is, the log-likelihood ratio admits a quadratic expansion in which the first-order term converges in distribution to a normal random variable with variance equal to twice the second-order term. This LAN condition is mild and holds for a broad class of smooth parametric models. For continuous $Y_j$, the LAN property follows from the \textit{differentiable in quadratic mean (DQM)} condition, which requires the existence of a measurable function $\score_{\pint}(Y_j|a,b)$ such that, as $\epsilon \to 0$, 
\begin{align*}
\int_{-\infty}^{+\infty}\left(\sqrt{\frac{\p_{\pint+\epsilon}(y|a,b)}{\p_{\pint}(y|a,b)}} - 1 - \frac{\epsilon}{2}\score_{\pint}(y|a,b)\right)^2\p_{\pint}(y|a,b)\dd y = o(\epsilon^2),
\end{align*}
for $a,b=1,\dots,K$. The DQM condition is implied by absolute continuity of the density function $\p_{\pint}(\cdot|a,b)$ with a square-integrable first-order derivative $\dot{\p}_{\pint}(\cdot|a,b)$. In that case, the score function is given by $\score_{\pint}(\cdot|a,b) = -\dot{\p}_{\pint}(\cdot|a,b)/\p_{\pint}(\cdot|a,b)$ and the Fisher information is $\FI_{\pint}(a,b) = \Exp_\pint\big[\score_{\pint}^2(Y_j|a,b)\big]$. For binary $Y_j$, the LAN condition arises from a standard second-order Taylor expansion, where the score function is $\score_{\pint}(Y|a,b) = Y - \pint_{a b}$ and the Fisher information is $\FI_{\pint}(a,b) = \pint_{a b}(1 - \pint_{a b})$; see \citet[Lemmas~1--2]{bickel2013asymptotic}.

Assumption~\ref{assm:LAN_individual}(b) can be heuristically motivated---though not formally implied---by part (a). Specifically, maximizing the LAN-form log-likelihood in (\ref{eqn:individual_LAN}) with respect to $\lpint$ yields its maximum likelihood estimate, $\hat\lpint_{a b}^\MLE = \FI_{\pint}^{-1}(a,b)\frac{1}{\sqrt{\m}}\sum_{j=1}^{\m}\score_{\pint}(Y_j|a,b)$ which weakly converges to $\mathcal{N}(0,\FI_{\pint}^{-1}(a,b))$; in addition, at $\pint_{a b}$, $\sqrt{\m}(\hat\pint_{a b}^\MLE - \pint_{a b}) = \hat\lpint_{a b}^\MLE$. We nonetheless impose part (b) as a formal assumption, which remains quite mild: For continuous $Y_j$, the asymptotic normality of the corresponding MLE is generally satisfied in smooth parametric models under mild regularity conditions; see, e.g., \citet[Chapter 5]{van2000asymptotic}. For binary $Y_j$, the asymptotic normality of the MLE follows directly from the central limit theorem for binomial experiments.

Define $\logiz_{a} = \log\frac{\pz(a)}{1 - \sum_{l=1}^{K-1}\pz(l)}$ for $a = 1,\dots,K-1$, and let $\logizb = (\logiz_{1},\dots,\logiz_{K-1})^\trans$. This one-to-one transformation allows us to use the notations $\f_{\pzb,\pintb}$, $\g_{\pzb,\pintb}$, and $\elbo_{\pzb,\pintb;\vq}$ interchangeably with $\f_{\logizb,\pintb}$, $\g_{\logizb,\pintb}$, and $\elbo_{\logizb,\pintb;\vq}$, respectively. The following proposition establishes the batched LAN result for the complete graph model in which $\vZ$ is observed.

\begin{proposition} \label{prop:LAN}
Assume Assumptions~\ref{assm:adaptive_policy} and~\ref{assm:LAN_individual} are satisfied. For all batch $t = 1,\dots,T$, the following results hold under $\f_{\logizb,\pintb}$. 
\begin{itemize}
\item[i)] The complete-graph log-likelihood ratio, $\lr_{\f,t}^{(\n,\m)}(\llogizb,\lpintb) \defeq \log\frac{\f_{\logizb+\llogizb/\sqrt{\n},\pintb+\lpintb/\sqrt{\m}}(\vY_t,\vZ_t)}{\f_{\logizb,\pintb}(\vY_t,\vZ_t)}$,  can be decomposed as
\begin{equation}
\begin{aligned}
\lr_{\f,t}^{(\n,\m)}(\llogizb,\lpintb)
=&~  \sum_{a=1}^{K-1}\llogiz_{a}\CS^{(\n)}_{\logizb,t}(a) - \frac{1}{2}\sum_{a=1}^{K-1}\sum_{b=1}^{K-1}\llogiz_{a}\llogiz_{b}\QV_{\logizb,t}(a,b) \\
 &~ + \sum_{a=1}^{K}\sum_{b=1}^{K}\left(\lpintab\CS^{(\m)}_{\pintb,t}(a,b) - \frac{1}{2}\lpintab^2\QV_{\pintb,t}(a,b)\right) + \op(1),  \\
\end{aligned}
\end{equation}
where
\begin{equation} \label{eqn:LAN_f_CSQV}
\begin{aligned}
\CS^{(\n)}_{\logizb,t}(a) &= \frac{1}{\sqrt{\n}}\sum_{i=1}^{\n}\left(\indicator\{\Z_{t,i} = a\} - \pz(a)\right),  \\
\QV_{\logizb,t}(a,b) &= 
      \begin{cases}
      \pz(a)(1-\pz(a)), & \text{if $a = b$} \\
      \pz(a)\pz(b), & \text{if $a\neq b$}  
      \end{cases} \\
\CS^{(\m)}_{\pintb,t}(a,b) &= \frac{1}{\sqrt{\m}}\sum_{j=1}^{\m} \indicator\{\Z_{t,i_1(j)} = a,\Z_{t,i_2(j)} = b\}\score_{\pint}(Y_{t,j}|a,b), \\
\QV_{\pintb,t}(a,b) &= \policy_t(a,b)\FI_{\pint}(a,b).
\end{aligned}
\end{equation}
\item[ii)] We have, as $\m_t\to\infty$ and $\n_t\to\infty$,
\begin{align} \label{eqn:LAN_f_CSconvergence}
\begin{pmatrix}
\CSb^{(\n)}_{\logizb,t} \\ \CSb^{(\m)}_{\pintb,t}
\end{pmatrix}
\wto
\begin{pmatrix}
\CSb_{\logizb,t} \\ \CSb_{\pintb,t}
\end{pmatrix} 
\sim
\mathcal{N}\left(
\begin{pmatrix}
\mfzero \\ \mfzero
\end{pmatrix}, 
\begin{pmatrix}
\QVb_{\logizb,t} & \mfzero \\ \mfzero & \QVb_{\pintb,t}
\end{pmatrix}\right),
\end{align}
where $\CSb^{(\n)}_{\logizb,t} = \big[\CS^{(\n)}_{\logizb,t}(a)\big]_{a=1}^{K-1}$, $\CSb^{(\m)}_{\pintb,t} = vech\big(\big[\CS^{(\m)}_{\pintb,t}(a,b)\big]_{a,b=1}^{K}\big)$, $\QVb_{\logizb,t} = \big[\QV_{\logizb,t}(a,b)\big]_{a,b=1}^{K-1}$, and $\QVb_{\pintb,t}  = \diag\big(\QV_{\pintb,t}(a,b)\big)_{a,b=1}^{K}$. Then,
\begin{equation*}
\begin{aligned}
\lr_{\f,t}^{(\n,\m)}(\llogizb,\lpintb) 
\wto
\lr_t(\llogizb,\lpintb) 
=&~ \sum_{a=1}^{K-1}\llogiz_{a}\CS_{\logizb,t}(a) - \frac{1}{2}\sum_{a=1}^{K-1}\sum_{b=1}^{K-1}\llogiz_{a}\llogiz_{b}\QV_{\logizb,t}(a,b) \\
 &~ + \sum_{a=1}^{K}\sum_{b=1}^{K}\left(\lpintab\CS_{\pintb,t}(a,b) - \frac{1}{2}\lpintab^2\QV_{\pintb,t}(a,b)\right).  \\
\end{aligned}
\end{equation*}
\item[iii)] Let $\big(\hat\logizb_{f,t}^\MLE,\hat\pintb_{f,t}^\MLE\big)$ denote the MLEs based on $\f_{\logizb,\pintb}(\vY_t,\vZ_t)$, we have 
\begin{align*}
\sqrt{\n}\big(\hat\logizb_{f,t}^\MLE - \logizb\big) &\sim \mathcal{N}\big(\mfzero, \QVb_{\logizb,t}^{-1}\big),  \\
\sqrt{\m}\big(\hat\pintb_{f,t}^\MLE - \pintb\big) &\sim \mathcal{N}\big(\mfzero, \QVb_{\pintb,t}^{-1}\big).
\end{align*}
\end{itemize} 
\end{proposition}

In establishing this result, the pre-specified policy condition in Assumption~\ref{assm:adaptive_policy} allows the pairing assignments $(i_{t,1}(j),i_{t,2}(j))^\trans$ (collected in $\vI_t$) to be treated as fixed, so the likelihood depends solely on the production outcomes $\Y_{t,j}$. The i.i.d.-ness of $\Y_{t,j}$ then implies that the joint likelihood of $\vY_t$ factorizes into the product of individual likelihoods, from which Proposition~\ref{prop:LAN} follows readily under the high-level conditions in Assumption~\ref{assm:LAN_individual}. A detailed proof is provided in Appendix~\ref{appendix:proofs} for completeness.

The complete graph LAN result in Proposition~\ref{prop:LAN}, developed at a given point $(\pzb,\pintb)$ in the parameter space, characterizes also the asymptotic behavior of statistics in its local neighborhood. Specifically, by Le Cam's third lemma, under the local alternative $\f_{\logizb+\frac{\llogizb}{\sqrt{\n}},\pintb+\frac{\lpintb}{\sqrt{\m}}}$ and as $\m_t\to\infty$ and $\n_t\to\infty$,
\begin{align} \label{eqn:LAN_f_CSconvergence_alternative}
\begin{pmatrix}
\CSb^{(\n)}_{\logizb,t} \\ \CSb^{(\m)}_{\pintb,t}
\end{pmatrix}
\wto
\begin{pmatrix}
\CSb_{\logizb,t} \\ \CSb_{\pintb,t}
\end{pmatrix} 
\sim
\mathcal{N}\left(
\begin{pmatrix}
\llogizb \\ \lpintb
\end{pmatrix}, 
\begin{pmatrix}
\QVb_{\logizb,t} & \mfzero \\ \mfzero & \QVb_{\pintb,t}
\end{pmatrix}\right),
\end{align}
for $\logizb\in\SR^{K}$ and $\lpintb\in\SR^{K^2}$.

More importantly, the LAN result within each batch invokes the \textit{Asymptotic Representation Theorem} (see \cite{le1972limits}, \cite{hajek1970characterization}, \cite{van2000asymptotic}). This theorem states that a sequence of complex experiments---here, the WSBM-based network production model---can be approximated by a simpler \textit{limit experiment}, whose likelihood ratio is the limit of the original sequence of likelihood ratios. In our context, the limit experiment corresponds to the so-called \textit{Gaussian shift experiment} in (\ref{eqn:LAN_f_CSconvergence_alternative}), where we observe single observations, or signals, of random variables $\QVb_{\logizb,t}^{-1}\CSb_{\logizb,t} \sim \mathcal{N}(\llogizb, \QVb_{\logizb,t}^{-1})$ and $\QVb_{\pintb,t}^{-1}\CSb_{\pintb,t} \sim \mathcal{N}(\lpintb, \QVb_{\pintb,t}^{-1})$. A direct calculation confirms that this limit experiment has a likelihood ratio equal to $\lr_t(\llogizb,\lpintb)$. 
% Additionally, the Asymptotic Representation Theorem states that for any convergent sequence of statistics from original experiments converges to a statistic in the limit experiment. This property allows us to first solve the inferential problem in the limit experiment and then translate the solution back to a finite-sample representation in the original experiments.

Across batches, the pre-specified policy condition (Assumption~\ref{assm:adaptive_policy}) and the batchwise application of variational estimation ensure that the sequence of batched likelihood ratios converges jointly: $$\big(\lr_{\f,1}^{(\n,\m)}(\llogizb,\lpintb),\dots,\lr_{\f,T}^{(\n,\m)}(\llogizb,\lpintb)\big) \wto \big(\lr_1(\llogizb,\lpintb),\dots,\lr_T(\llogizb,\lpintb)\big).$$ This invokes the multi-stage Asymptotic Representation Theorem (Theorem~2 of \cite{hirano2023asymptotic}), which states that any statistical decision rule based on a convergent sequence of statistics in the original batched network formation problem admits a corresponding representative rule in the limiting $T$-observation Gaussian shift experiments. We will leverage this result to develop our adaptive network formation policy in Section~\ref{sec:onlinnetworkformation_detail}.

It is worth noting that the Gaussian signals across batches are not identically distributed. In particular, the normal distribution $\mathcal{N}(\lpintb, \QVb_{\pintb,t}^{-1})$ has batch-specific variance determined by the network formation policy $\policy_t$ and the production model (via Fisher information $\FI_{\pint}(a,b)$). As shown in (\ref{eqn:LAN_f_CSQV}), the variance is given by $\QV_{\pintb,t}^{-1}(a,b) = \frac{1}{\policy_t(a,b)\FI_{\pint}(a,b)}$. This reflects the intuition that, for fixed $\FI_{\pint}(a,b)$, assigning more $(a,b)$-type pairs (i.e., a larger $\policy_t(a,b)$ value) leads to more precise estimation of the production parameter $\pint_{a b}$, hence lower variance.

\subsection{Asymptotic Results under Unobserved Agent Types} \label{subsec:LAN_Z_unobserved}
We now return to the real-world setting where the decision maker cannot directly observe the agent types $\vZ_t$. We show below that the marginal model $\g_{\pzb,\pintb}(\vY_t)$, and subsequently its variational approximation $\elbo_{\logizb,\pintb;\vq}(\vY_t)$, is asymptotically equivalent to the complete graph model $\f_{\pzb,\pintb}(\vY_t,\vZ_t)$. 

This result is developed under a \textit{batched sparse network asymptotic} regime, formalized in Assumption~\ref{assm:asymptotic_sparsewithinbatch}. This mirrors the sparse network asymptotics for the classical stochastic block model (see, e.g., \cite{bickel2013asymptotic}, \cite{amini2013pseudo}), where the expected number of edges per node, or \textit{degree}, grows at the rate $O(\log\n_t)$. Such a growth rate reflects empirical patterns in large-scale networks, where each node typically forms only a few links. Compared to \textit{dense network} asymptotics, the sparse regime yields more realistic and accurate approximations in these dateset. In our batched WSBM setting, we similarly allow the average the number of (pre-specified) edges per node, $\m_t/\n_t$, in each batch $t$ to grow as slowly as $O(\log\n_t)$ as the number of nodes $\n_t$ increases. This is also motivated by real-world scenarios, such as the sales-force context in \cite{xu2024heterogeneous}, where each agent can only participate in a limited number of tasks within a given batch period (e.g., one month), regardless of the total number of potential collaborators. We formalize this condition as follows. 

\begin{assumption}[Batched Sparse Network] \label{assm:asymptotic_sparsewithinbatch}
For each batch $t = 1,\dots,T$, we let $\m_t \to \infty$ as $\n_t\to\infty$ such that $\m_t/(\n_t\log\n_t) \to \rate_t$ for some $\rate_t \in (0,\infty)$. 
\end{assumption}

To establish the desired equivalence result, we also impose additional regularity conditions on the production model, given in Assumptions~\ref{assm_BickelChen2009} and~\ref{assm:likelihoodmodularity}. These conditions are readily verified once the distribution is specified. For instance, in the case of continuous $Y_j$, they are satisfied when $\f$ is normal. In the case of binary outcomes $Y_j$, these conditions, adapted to our weighted SBM setting, can likewise be verified following the approach of \cite{bickel2013asymptotic} for proving their Theorem~1. These conditions, together with the proof, are organized in Appendix~\ref{appendix:proofs}.

\begin{theorem} \label{thm:fg_equal}
Assume that $\f_{\pzb,\pintb}$ is unimodal in $(\pzb,\pintb)$, that $(\pint_{1 b},\dots,\pint_{K b})^\trans \neq (\pint_{1 b'},\dots,\pint_{K b'})^\trans$ for all $b \neq b' \in \{1,\dots,K\}$, and that Assumptions~\ref{assm:adaptive_policy}--\ref{assm:likelihoodmodularity} are satisfied. Let $\mathcal{S}(\pzb,\pintb)$ denote the equivalent set of $(\pzb,\pintb)$ up to a relabeling of the latent classes $\{1,\dots,K\}$. 
\begin{itemize}
\item[(i)]  Then, we have
\begin{align}
\g_{\pzb,\pintb}(\vY_t) = \max_{(\pzb^\prime,\pintb^\prime)\in\mathcal{S}(\pzb,\pintb)}\f_{\pzb^\prime,\pintb^\prime}(\vY_t,\vZ_t) + o_P(1),
\end{align}
where the $o_P(1)$ term is uniform over $\pzb\in\SR^{K}$ and $\pintb\in\SR^{K^2}$.
\item[(ii)] Let $(\hat\logizb_{g,t}^\MLE,\hat\pintb_{g,t}^\MLE) = \argmax_{\logizb\in\SR^{K},\pintb\in\SR^{K^2}}\g_{\logizb,\pintb}(\vY_t)$ be the MLEs, then its relabeling equivalence class $\mathcal{S}(\hat\logizb_{g,t}^\MLE,\hat\pintb_{g,t}^\MLE)$ contains an element $(\hat\logizb_{g,t}^\prime,\hat\pintb_{g,t}^\prime)$ such that
\begin{align*}
\sqrt{\n}(\hat\logizb_{g,t}^\prime - \logizb) &\sim \mathcal{N}\big(\mfzero, \QVb_{\logizb,t}^{-1}\big),  \\
\sqrt{\m}(\hat\pintb_{g,t}^\prime - \pintb) &\sim \mathcal{N}\big(\mfzero, \QVb_{\pintb,t}^{-1}\big).
\end{align*}
\end{itemize}
\end{theorem}

Theorem~\ref{thm:fg_equal}(i) shows that, under batched sparse network asymptotics, the likelihoods with and without observing the latent agent types $\vZ_t$ are asymptotically equivalent. This, in turn, implies that the MLEs based on the complete-graph likelihood $\f(\vY_t,\vZ_t)$ and those based on the marginal likelihood $\g(\vY_t)$ are asymptotically equivalent, as formalized in part (ii). While this result primarily serves as a theoretical intermediate step, since maximizing $\g(\vY_t)$ is generally infeasible due to the computational challenges discussed in Section~\ref{sec:estimation}, it nonetheless provides a critical foundation for analyzing the variational approximation $\elbo(\vY_t)$. In Theorem~\ref{thm:variational_LAN} below, we establish that $\elbo(\vY_t)$ is asymptotically equivalent to $\log\f(\vY_t,\vZ_t)$ under the same sparse asymptotics. As a result, feasible variational estimates based on $\elbo(\vY_t)$ are asymptotically equivalent to the oracle MLEs based on the complete-data likelihood. The proof is provided in Appendix~\ref{appendix:proofs}.

\begin{theorem} \label{thm:variational_LAN}
Assume the same conditions as in Theorem~\ref{thm:fg_equal} hold.
\begin{itemize}
\item[(i)] Define $\bar\elbo_{\logizb,\pintb}(\vY_t) = \max_{\vq\in\SQ} \elbo_{\logizb,\pintb;\vq}(\vY_t)$, we have 
\begin{align*}
\bar\elbo_{\logizb,\pintb}(\vY_t) = \max_{(\logizb^\prime,\pintb^\prime)\in\mathcal{S}(\pzb,\pintb)}\log\f_{\logizb^\prime,\pintb^\prime}(\vY_t,\vZ_t) + o_P(1),
\end{align*}
where the $o_P(1)$ term is uniform over $\pzb\in\SR^{K}$ and $\pintb\in\SR^{K^2}$.
\item[(ii)] Let $(\hat\logizb_{\elbo,t}^\VE,\hat\pintb_{\elbo,t}^\VE) = \argmax_{\logizb\in\SR^{K},\pintb\in\SR^{K^2}} \bar\elbo_{\logizb,\pintb}(\vY_t)$ be the mean-field variational estimates. Then, its relabeling equivalence class $\mathcal{S}(\hat\logizb_{\elbo,t}^\VE,\hat\pintb_{\elbo,t}^\VE)$ contains an element $(\hat\logizb_{\elbo,t}^\prime,\hat\pintb_{\elbo,t}^\prime)$ such that
\begin{align*}
\sqrt{\n}(\hat\logizb_{\elbo,t}^\prime - \logizb) &\sim \mathcal{N}\big(\mfzero, \QVb_{\logizb,t}^{-1}\big),  \\
\sqrt{\m}(\hat\pintb_{\elbo,t}^\prime - \pintb) &\sim \mathcal{N}\big(\mfzero, \QVb_{\pintb,t}^{-1}\big).
\end{align*}
\end{itemize} 
\end{theorem}

Theorem~\ref{thm:variational_LAN}(i) further implies that the $\elbo_{\logizb,\pintb;\vq}(\vY)$-based log-likelihood ratio proxy, $\lr_{\elbo,t}^{(\n,\m)}(\llogizb,\lpintb,\vq) = \frac{\elbo_{\logizb+\llogizb/\sqrt{\n},\pintb + \lpintb/\sqrt{\m};\vq}(\vY)}{\elbo_{\logizb,\pintb;\vq}(\vY)}$, is asymptotically equivalent to the complete-graph log-likelihood ratio, $\lr_{\f,t}^{(\n,\m)}(\llogizb,\lpintb)$; both converge weakly to the same Gaussian shift experiment likelihood ratio $\lr_t(\llogizb,\lpintb)$ (defined in Proposition~\ref{prop:LAN}). Using arguments similar to those in Proposition~\ref{prop:LAN}, we obtain a LAN-type expansion for $\lr_{\elbo,t}^{(\n,\m)}(\llogizb,\lpintb,\vq)$ as
\begin{equation*}
\begin{aligned}
\lr_{\elbo,t}^{(\n,\m)}(\llogizb,\lpintb,\vq) 
=&~ \sum_{a=1}^{K-1}\llogiz_{a}\CS^{(\n)}_{\elbo,\logizb,t}(a) - \frac{1}{2}\sum_{a=1}^{K-1}\sum_{b=1}^{K-1}\llogiz_{a}\QV_{\elbo,\logizb,t}(a,b)\llogiz_{b} \\
&~ + \sum_{a=1}^{K}\sum_{b=1}^{K}\left(\lpintab\CS^{(\m)}_{\elbo,\pintb,t}(a,b) - \frac{1}{2}\lpintab^2\QV_{\elbo,\pintb,t}(a,b)\right) + o_P(1),
\end{aligned}
\end{equation*}
where
\begin{align*}
\CS^{(\n)}_{\elbo,\logizb,t}(a) &= \frac{1}{\sqrt{\n}}\sum_{i=1}^{\n}\left(q_i(a) - \pz(a)\right),  \\
\QV_{\elbo,\logizb,t}(a,b) &= 
      \begin{cases}
      \pz(a)(1-\pz(a)), & \text{if $a = b$} \\
      \pz(a)\pz(b),       & \text{if $a\neq b$}  
      \end{cases} \\
\CS^{(\m)}_{\elbo,\pintb,t}(a,b) &= \frac{1}{\sqrt{\m}}\sum_{j=1}^{\m} \qjl(a)\qjr(b)\score_{\pint}(Y_j|a,b), \\
\QV_{\elbo,\pintb,t}(a,b) &= \frac{1}{\m}\sum_{j=1}^{\m} \qjl(a)\qjr(b)\FI_{\pint}(a,b).
\end{align*}
It is straightforward to verify that $\lr_{\elbo,t}^{(\n,\m)}(\llogizb,\lpintb,\vZ) = \lr_{\f,t}^{(\n,\m)}(\llogizb,\lpintb)$, meaning the ELBO function recovers the complete-graph log-likelihood ratio when the variational approximations $q_i(a)$ in the former are replaced by the true class indicators $\indicator\{\Z_{t,i} = a\}$. Therefore, their asymptotic equivalence implies that each $q_i(a)$ converges to $\indicator\{\Z_{t,i} = a\}$ within each batch $t$ under the sparse network asymptotics. This aligns with the established consistency results for label estimation of $\vZ$ in the classical stochastic block models using methods such as the EM algorithm; See, e.g., \cite{bickel2009nonparametric}, \cite{amini2013pseudo}, \cite{wang2017likelihood} for more detailed results. That said, our online network formation algorithm does not aim to explicitly recover the latent labels $\vZ$ within each batch. Instead, it leverages $\vq$ in its capacity as batched categorical approximations to the posterior distribution of $\vZ$, as elaborated in the next section. 

%Define $\lr_{\elbo}^{(\n,\m)}(\llogizb,\lpintb) \defeq \bar\elbo_{\logizb+\frac{\llogizb}{\sqrt{\n}},\pintb+\frac{\lpintb}{\sqrt{\m}}}(\vY) - \bar\elbo_{\logizb,\pintb}(\vY)$, where . We have, for each batch,
%\begin{align*}
%\lr_{\elbo}^{(\n,\m)}(\llogizb,\lpintb) \wto \lr(\llogizb,\lpintb).
%\end{align*} 

%Moreover, the variational estimates $\hat\pzb$ and $\hat\pintb$, obtained via the EM algorithm in Algorithm~\ref{algorithm:EM}, are consistent and asymptotically normal.

%Denote by $\hat{\vZ}$ the maximizer of $\modula^{(\n,\m)}(\vY,\ve)$. Then, up to a permutation $\tau$ on $\{1,\dots,K\}$, we have 
%\begin{align*}
%\prob(\hat{\vZ} \neq \tau(\vZ)) \to O(n^{-b_n})
%\end{align*}
%for some sequence $b_n \to \infty$. 

%\begin{remark}
%The likelihood of the CGM model associate with latent type vector $\ve$ is similarly given by
%\begin{align*}
%\f_{\pzb,\pintb}(\vY,\ve) = \left(\prod_{i=1}^{\n}\pz(\e_i)\right)\prod_{j=1}^{\m} \p_{\pint}\big(Y_j|\e_{i_1(j)},\e_{i_2(j)}\big)
%\end{align*}
%Let $Q_n(\vY,\ve) = \sup_{\pintb}\log\f_{\pzb,\pintb}(\vY,\ve)$, the likelihood modularity by \cite{bickel2009nonparametric}, which equals
%\begin{align*}
%Q_n(\vY,\ve) =&~ \sum_{a = 1}^{K} \n_a\log\frac{\n_a}{\n} + \frac{1}{2}\sum_{a=1}^{K}\sum_{b=1}^{K}\left(O_{a b}\log\frac{O_{a b}}{\m_{a b}} + \left(\m_{a b} - \frac{O_{a b}}{\m_{a b}}\right)\log\left(1 - \frac{O_{a b}}{\m_{a b}}\right)\right)
%\end{align*}
%\end{remark}

\section{Online Network Formation Algorithm} \label{sec:onlinnetworkformation_detail}
This section details the implementation of our online network formation algorithm. Section~\ref{subsec:bayesian_update} describes the Bayesian updating procedure, which refines posterior beliefs about the pairwise production parameters and latent agent types using batched variational estimates as signals. In Section~\ref{subsec:maximumweightmatching}, we incorporate these updated posteriors and translate the policy-level constrained optimization problem from Section~\ref{sec:onlinenetworkformation} to the agent level, making it readily applicable in practice. Finally, Section~\ref{subsec:algorithms} presents the complete algorithmic procedure.

\subsection{Bayesian updating and classification} \label{subsec:bayesian_update}
The theoretical results in Section~\ref{sec:asmptotic_results} convey a simple yet powerful insight: the complex batched network formation and production problem can be approximated by a multi-stage Gaussian shift experiment. Specifically, for pairwise production parameter $\pint_{a b}$, where $a, b = 1, \dots, K$, we receive a single Gaussian signal at each stage $t = 1,\dots,T$, with mean $\pintb$ and variance determined by the network formation policy $\policy_t(a,b)$ and the production model $\p_{\pint}(\cdot|a,b)$. Under Gaussianity, these signals are thus stage-specific maximum likelihood estimates (MLEs) for $\pintb$. Since the variational estimates are shown to converge weakly to these MLEs, they serve as feasible finite-sample analogues.

In light of this, we apply Gaussian Bayesian updating pairwise to each $\pint_{a b}$ using their batched variational estimates. For each $(a,b)$ pair with $a,b = 1,\dots,K$, let $\hat\pint_{t,a b}$ denote the variational estimate of $\pint_{a b}$ from batch $t$, and let $\widehat\se_{t,a b}$ denote its standard error. Assuming a Gaussian prior $\mathcal{N}(\mu_{0,a b}, \sigma_{0,a b}^2)$, the posterior distribution after round $t$ remains Gaussian, $\mathcal{N}(\mu_{t,a b}, \sigma_{t,a b}^2)$, and is updated recursively via the standard Gaussian Bayesian formula
\begin{equation} \label{eqn:Bayesianupdating_pintb}
\begin{aligned}
\mu_{t,a b} &= \bigg(\frac{\mu_{t-1,a b}}{\sigma_{t-1,a b}^2} + \frac{\hat\pint_{t,a b}}{\widehat\se_{t,a b}^2}\bigg)\bigg/\bigg(\frac{1}{\sigma_{t-1,a b}^2} + \frac{1}{\widehat\se_{t,a b}^2}\bigg), \\
\sigma_{t,a b}^2 &= 1\big/\bigg(\frac{1}{\sigma_{t-1,a b}^2} + \frac{1}{\widehat\se_{t,a b}^2}\bigg),
\end{aligned}
\end{equation}
for $t = 1,\dots, T$ and $a,b = 1,\dots,K$. 

Turning to the agents' latent types, recall that for each agent $i$, the variational estimation via EM algorithm in each batch $t$ produces a local variational approximation $\hat\q_{t,i}(a)$, a categorical distribution over types $a = 1,\dots,K$, approximating the posterior of the agent's type $\Z_i$. These categorical signals can be aggregated recursively for each agent via Bayesian classification. Assuming a categorical prior $\omega_{0,i}(a)$ on $\Z_i$, the posterior after batch $t$, denoted by $\omega_{t,i}(a)$, remains categorical and updates as
\begin{align} \label{eqn:Bayesianupdating_vq}
\omega_{t,i}(a) = \frac{\omega_{t-1,i}(a)\hat{q}_{t,i}(a)}{\sum_{a=1}^{K}\omega_{t-1,i}(a)\hat{q}_{t,i}(a)},
\end{align}
for $t = 1,\dots,T$, $i = 1,\dots,\n$, and $a = 1,\dots,K$. 

Beyond a natural choice for sequential learning, Bayesian updating offers an additional practical advantage: it systematically incorporates prior knowledge about unknown parameters, particularly valuable in real-world applications. For instance, managers or teachers may have prior insights into workers’ styles or students’ personalities when forming teams or assigning seatmates. Similarly, users on social platforms often disclose latent traits during registration. When no prior information is available, the framework defaults to uninformative priors: infinite-variance normal distributions for $\pint_{a b}$ and discrete uniform distributions for $Z_i$.

\subsection{Maximum weight matching} \label{subsec:maximumweightmatching}
In practical settings, the policy-level constrained optimization problem in (\ref{eqn:optimization_oracle_constrained}) must often be reformulated at the agent level for (at least) two key reasons. First, certain practical constraints need to be implemented at the individual level. For instance, the workload constraint reflects the fact that each agent can handle only a limited number of tasks within a given time frame and must therefore be applied individually. Second, reliable inference of latent types requires that each agent be matched frequently enough to ensure the convergence of their estimated type distribution $\hat\q_i(a)$). We now reformulate the problem accordingly as a agent-level \textit{maximum weight matching} problem.

Let $\post\pint_{a b}$ and $\post\Z_{i}$ denote some realizations (e.g., point estimates, random draws) of the pairwise production parameters and agent latent types, respectively, obtained from their latest posteriors. Based on them, we aim to solve
\begin{align} \label{eqn:optimization_individuallevel}
\max_{\substack{\x_{i i'}\in\{0,1\}}} \, \sum_{i=1}^{\n}\sum_{i'=i+1}^{\n} \x_{i i'} \, \post\outfun(\post\Z_{i},\post\Z_{i'})
\end{align}
subject to 
\begin{align} \label{eqn:constraint_m_individuallevel}
\sum_{i=1}^{\n}\sum_{i'=i+1}^{\n} \x_{i i'} = \m,
\end{align}
where $\post\outfun(a,b) = \int_{y\in\SY} \p_{\post\pint}(y|a,b)\dd y$ is the expected weight for a type $a$-$b$ pair, and $\x_{i i'} = 1$ indicates that agents $i$ and $i'$ are paired (otherwise, $\x_{i i'} = 0$). In words, by selecting which edges to form among all possible agent pairs, we seek to maximize the overall expected output (\ref{eqn:optimization_individuallevel}), subject to the constraint (\ref{eqn:constraint_m_individuallevel}) that exactly $\m$ edges are formed. 

We next reformulate the workload and clipping constraints in (\ref{eqn:constraint_workload}) and (\ref{eqn:constraint_clipping}) at the agent level. The workload constraint becomes
\begin{align} \label{eqn:constraint_workload_individuallevel}
\sum_{i'=1}^{i-1} \x_{i' i} + \sum_{i'=i+1}^{\n} \x_{i i'} \in [\bound_l, \bound_h]
\end{align}
for some $\bound_l \leq \bound_h \in \SN^{+}$, ensuring that each agent is assigned a number of tasks within the allowed range. The clipping constraint is similarly recast as
\begin{align} \label{eqn:constraint_clipping_individuallevel}
\sum_{i=1}^{\n}\sum_{i'=i+1}^{\n} \x_{i i'}\indicator\{\post\Z_{t,i}=a, \post\Z_{t,i'}=b\} \in [\lceil\m\rate\rceil,\,\lfloor\m(1-\rate)\rfloor], 
\end{align}
for a given clipping rate $\rate\in(0, 0.5)$ and all $a,b = 1,\dots,K$.

The maximum weight matching problem in (\ref{eqn:optimization_individuallevel}), subject to (\ref{eqn:constraint_m_individuallevel}) and, when relevant, (\ref{eqn:constraint_workload_individuallevel}) and (\ref{eqn:constraint_clipping_individuallevel}), can be formulated as an integer linear program, the solution to which yields a valid matching that maximizes total weight.

\subsection{HGT algorithm} \label{subsec:algorithms}
In the classical MAB problem, the central challenge is balancing exploration (pulling less-explored, potentially suboptimal arms to gain information) and exploitation (pulling arms currently believed to yield the highest reward). This trade-off is typically managed by policies that determine arm-pulling probabilities, which can often be interpreted as functions of the posteriors of the unknown rewards. In particular, the \textit{greedy} algorithm pulls the arm with the highest average reward, corresponding to the point estimate provided by the Gaussian posterior means. \textit{Thompson sampling} (\citet{thompson1933likelihood}) assigns probabilities based on the chance each arm is optimal: it draws a sample from each arm’s posterior and selects the arm with the highest draw. The \textit{Upper Confidence Bound (UCB)} algorithm (see, e.g., \cite{lai1985asymptotically}, \cite{auer2002using}) chooses the arm with the highest upper bound of a confidence (or credible, under Gaussianity) interval, thereby allocating more probability to less-explored arms, which tend to have wider bounds.

Leveraging these insights, we construct our online network formation algorithm using the posteriors $\mathcal{N}(\mu_{t,a b}, \sigma_{t,a b}^2)$ for $\pint_{a b}$ and $\omega_i$ for $Z_i$. Specifically, we sample agent-type realizations $\tilde\Z_i$ from categorical posteriors $\omega_i$, following the principle of Thompson sampling. Given the exploration inherent in sampling $\tilde\Z_i$, we we adopt a greedy heuristic for $\pint_{a b}$ by simply using the posterior mean as a point estimate, setting $\tilde\pint_{a b} = \mu_{t,a b}$. These sampled types and point estimates, denoted with tildes, are then used in the maximum weight matching optimization described in Section~\ref{subsec:maximumweightmatching}, whose solution defines the sampling policy for the next round. We refer to this ``greedy-in-$\pintb$, Thompson-in-$\vq$'' strategy as the \textit{Hybrid Greedy-Thompson (HGT) algorithm}, detailed in Algorithm~\ref{algorithm:HGT}.

\medskip
\begin{algorithm}[H]
\caption{Hybrid Greedy-Thompson (HGT) Algorithm} \label{algorithm:HGT}
\vspace{0.9em}  % Adds some negative space between the caption and the rule

\textbf{Require:} Number of batches $T$; Number of types $K$; Priors for $\pint_{a b}$: $\mathcal{N}(\mu_{0,a b},\sigma_{0,a b}^2)$, $a, b = 1,\dots,K$; Priors for $\Z_i$: $categorical(\omega_{0,i}(1),\dots,\omega_{0,i}(K))$, $i=1,\dots,n$.

% \vspace{0.5em}
% \noindent \textit{Initialize:} Assign random pairs for batch \(t=1\) subject to constraints.

\vspace{0.9em}
\noindent Iterate the following steps for batch \(t = 1, \dots, T\):
\begin{itemize}
    \item[-] \textit{Step 1: Parameter Realization.} 
    For each pairwise production parameter $\pint_{a b} \sim \mathcal{N}(\mu_{t-1,a b},\sigma_{t-1,a b}^2)$, set $\tilde\pint_{a b} = \mu_{t-1,ab}$; 
    For each agent's latent type $\Z_i$, draw a realization $\tilde\Z_i$ from $categorical(\omega_{t-1,i}(1),\dots,\omega_{t-1,i}(K))$.
    
    \item[-] \textit{Step 2: Network Formation and Production.} Solve the maximum-weight matching problem in (\ref{eqn:optimization_individuallevel}) subject to constraints (\ref{eqn:constraint_m_individuallevel}), and (potentially) also to (\ref{eqn:constraint_workload_individuallevel}) and (\ref{eqn:constraint_clipping_individuallevel}), to determine the agent-level pairing policy; Record the network formation outcome in $\vI_t$ and the production outcome in $\vY_t$.

    \item[-] \textit{Step 3: Variational Estimation.} Using $(\vI_t,\vY_t)$, compute variational estimates $\hat{\theta}_{t,a b}$ and $\hat{q}_{t,i}$ via the Expectation-Maximization algorithm described in Section~\ref{sec:estimation}.

    \item[-] \textit{Step 4: Bayesian Updating.} Update the posterior distributions of production parameters, $\theta_{a b}$, and agent types, $\Z_i$, using \eqref{eqn:Bayesianupdating_pintb} and \eqref{eqn:Bayesianupdating_vq}, respectively.
\end{itemize}
\end{algorithm}
\medskip

\section{Simulation} \label{sec:simulation}
% [old version]
% In this section, we carry out a comprehensive Monte Carlo study to assess the performance of the Hybrid-Greedy-Thompson (HGT) algorithm proposed in Section~\ref{sec:onlinnetworkformation_detail} for online network formation. We simulate an environment in which a decision maker assigns agents to pairs to complete tasks with binary outcomes, i.e., $Y_j \in \{0, 1\}$. We explore two settings separately, where agents are grouped into either two types ($K = 2$) or three types ($K = 3$), each with distinct model specifications. All results are based on $100$ repetitions. Since the likelihood in the WSBM is often susceptible to multiple local optima (see, e.g., \cite{nowicki2001estimation}, \cite{bickel2013asymptotic}, \cite{bonhomme2021teams}), we initialize the variational EM algorithm from multiple starting points.

We conduct a series of Monte Carlo experiments to evaluate the finite-sample performance of the Hybrid-Greedy-Thompson (HGT) algorithm introduced in Section~\ref{sec:onlinnetworkformation_detail}. In each experiment, HGT assigns agents into pairwise networks to complete binary-outcome tasks ($Y_j \in \{0, 1\}$). We consider two cases: one with two latent agent types ($K = 2$) and the other with three ($K = 3$), which allows us to assess the algorithm’s robustness to different levels of latent structure. In both cases, key parameters are chosen to broadly reflect real-world settings commonly studied in the empirical literature.

For the $K = 2$ case, we draw on the classroom seating and friendship formation context in \cite{rohrer2021proximity} to guide our choices of network size and the productivity regime. That study finds that students with similar observable characteristics tend to form friendships more successfully. While our model focuses on unobserved heterogeneity (latent types), we retain the underlying intuition by letting agents with similar latent types have a higher probability of productive interaction. In the $K = 3$ case, we calibrate the agent-to-task ratio and average productivity level to be aligned with \cite{xu2024heterogeneous}, which studies salesforce team formation and success in real estate deals. A key finding there is that agents with intermediate solo performance tend to have better team productivity when paired with any other type of agent. We retain this feature in our data-generating process by assigning one latent type to consistently deliver strong team performance across all pairings.
% Although our model specifications differ from these empirical applications, calibrating production parameter values to empirically relevant ranges helps make the simulated environments more realistic and policy-relevant.

For the implementation of the EM algorithm, we initialize from multiple random starting points to reduce sensitivity to local optima, a widely recognized issue for SBM-based models (see, e.g., \cite{nowicki2001estimation}, \cite{bickel2013asymptotic}, \cite{bonhomme2021teams}). All simulation results are based on $100$ repetitions.

\subsection{Case of $K = 2$}
% We begin with the K = 2 case, which corresponds to a simplified educational setting inspired by the classroom seating experiment in \cite{rohrer2021proximity}. 
We consider a fixed pool of $n = 32$ agents (e.g., students), slightly smaller than the maximum class size studied in \cite{rohrer2021proximity}. Each student belongs to one of $K = 2$ latent types, representing unobserved characteristics such as personality traits that are unlikely to change over short time horizons. In each replication, student types are independently drawn from a multinomial distribution with equal probabilities $(0.5, 0.5)$. Once generated, these types remain fixed throughout the simulation and are treated as unknown parameters. The probability of $Y_j = 1$ for a given pair $j$---indicating friendship formation---is specified by:
\begin{align*}
\pintb
=
\begin{pmatrix}
0.18 & 0.13 \\
0.13 & 0.50
\end{pmatrix},
\end{align*}
with rows and columns ordered by type $(1, 2)$. 
% This setup implies that one type is generally more likely to form a social tie, and that same-type pairs tend to achieve higher success rates.

We run our HGT algorithm over $T = 6$ batches. In each batch, the $32$ students are assigned to $384$ seatmate pairs, resulting in an average of $24$ ($= 384\times2/32$) pairings per student. This corresponds to a teacher who reshuffles student seating on a fixed schedule (e.g., weekly), collects data, and updates the pairing policy after each batch. Depending on the frequency of reshuffling, each batch can be interpreted as spanning several months or one or more academic semesters.
% \footnote{For example, if seating is updated every week, one batch would correspond to roughly 1.5 semesters; if updated every two weeks, one batch would span about three semesters.} 
Since each student participates in exactly the same number of pairings per batch, we impose the agent-level ``workload'' constraint in (\ref{eqn:constraint_workload_individuallevel}) with $\bound_l = \bound_h = 24$ in this case. 

\begin{figure}[H]
    \centering
    \bigskip{}
    \includegraphics[scale=0.7]{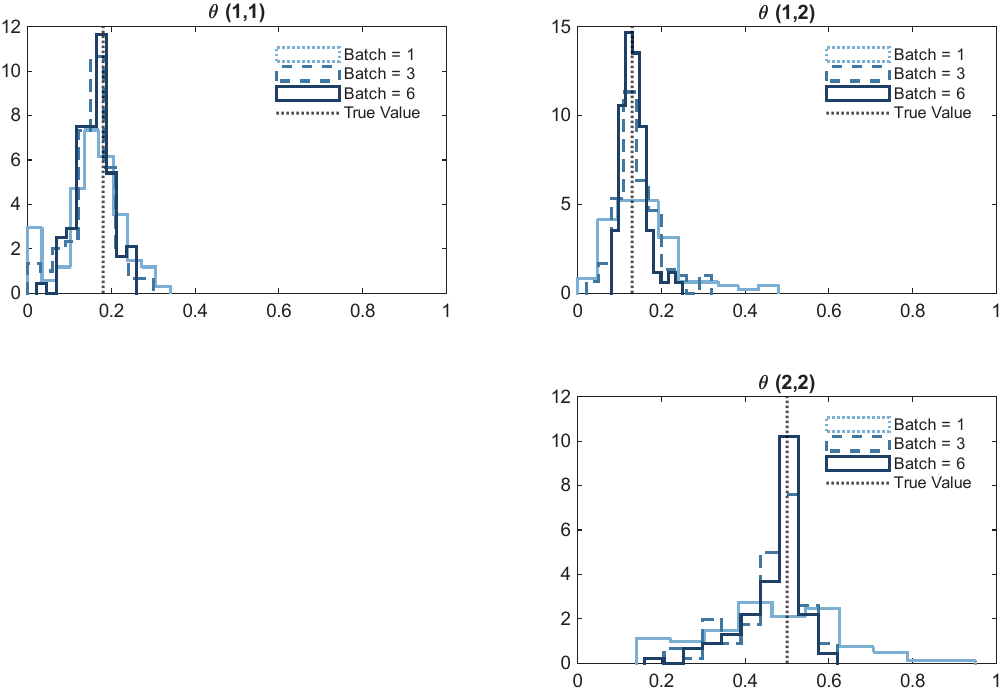}
    \medskip{}
    
    \begin{minipage}{1\textwidth} % Set width equal to the figure width
        \caption{\footnotesize Histograms of empirical posteriors for $\pintb$ based on batched variational estimates after batch $1$ (light blue), $3$ (blue), and $6$ (dark blue).}
        \label{fig:K2_empirical_posterior}
    \end{minipage}
\end{figure}

Figure~\ref{fig:K2_empirical_posterior} shows the empirical posterior distributions of each $\pint_{ab}$ for $a, b = 1, 2$, after batches 1, 3, and 6. The histograms become progressively more concentrated and shift toward the true parameter values, marked by grey dotted vertical lines. This indicates that the algorithm effectively refines the parameter estimates via Bayesian updating, though the estimates from each batch (appear to) exhibit small finite-sample biases (which may be attributable to the presence of multiple local optima discussed earlier).\footnote{Unreported simulation results suggest that these biases diminish as the sample size per batch increases.}

Table~\ref{tab:K2_falselabelingrates} reports the false labeling rates, defined as the proportion of incorrectly estimated $Z_i$ values across all $\n$ agents: $\sum_{i=1}^{\n} \indicator\{\hat{Z}_i \neq Z_i\}/\n$, where $\hat{Z}_i$ denotes the estimated latent type for agent $i$. Each estimate is assigned using the \textit{maximum a posteriori} rule: $\hat{Z}_i = \arg\max_{k} \tilde{q}_i(k)$, where $\tilde{q}_i$ is the posterior distribution after batch $t$. 
%Here, each $Z_i$ is estimated by $\hat{Z}_i = \argmax_{k} \tilde{q}_i(k)$, where $\tilde{q}_i$ is the posterior distribution obtained after batch $t$. 
From Table~\ref{tab:K2_falselabelingrates}, we observe that the false labeling rate declines steadily over successive batches, falling from nearly $22\%$ after the first batch to about $8\%$ the sixth. The steady decline in misclassification rates suggests that the HGT algorithm effectively accumulates information on each agent’s type over time through Bayesian updating based on batched LVAs, echoing the asymptotic insight in Section~\ref{sec:asmptotic_results} that latent type recovery improves as more data accumulate.

\begin{table}[H]
\centering
\begin{tabular}{lcccccc}
\toprule
Batch & 1 & 2 & 3 & 4 & 5 & 6 \\
\midrule
$\text{FLR (\%)}$ & 22.06 & 13.63 & 10.88 & 8.53 & 8.03 & 7.97 \\
\bottomrule
\end{tabular}
\caption{\small False Labeling Rates (FLR) across batches, reported in percentages.}
\label{tab:K2_falselabelingrates}
\end{table}

\begin{figure}[H]
    \centering
    \begin{minipage}[t]{0.45\textwidth}
        \centering
        \includegraphics[width=\linewidth]{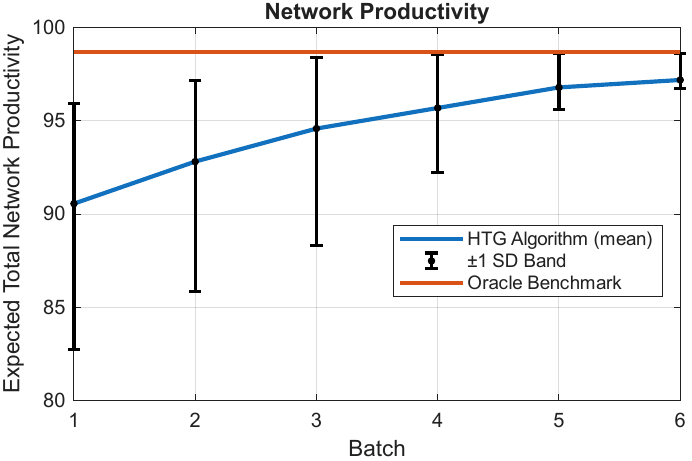}
    \end{minipage}%
    \hspace{0.02\textwidth}
    \begin{minipage}[t]{0.45\textwidth}
        \centering
        \includegraphics[width=\linewidth]{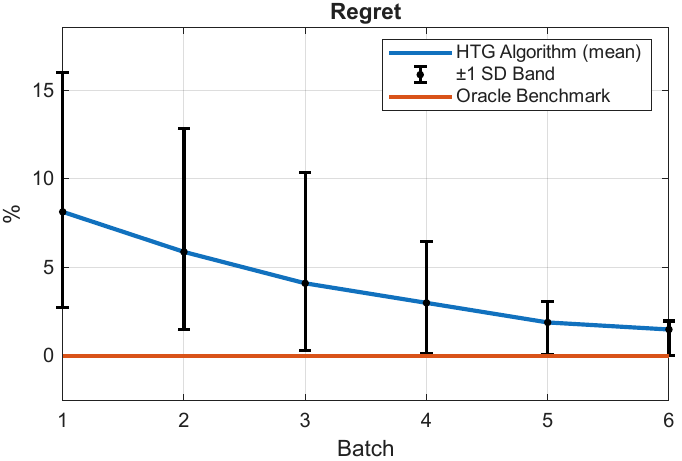}
    \end{minipage}
    %\caption{\footnotesize Network production performance of the Hybrid Greedy-Thompson (HGT) algorithm (blue), with error bars indicating $\pm 1$ standard deviation (black), compared to the oracle benchmark (orange). Performance is evaluated by expected total success rate (productivity; left) and percentage regret (right).}
    \caption{\small Network production performance of the Hybrid Greedy-Thompson (HGT) algorithm (blue) versus the oracle benchmark (orange). The left panel reports the average total success rate (network productivity), with error bars representing $\pm 1$ standard deviation (black) over 100 repetitions. The right panel displays regret, defined as the percentage gap in expected network output relative to the oracle.}
    \label{fig:K2_networkproduction}
\end{figure}

We now turn to network production performance---the primary economic outcome of interest. The left panel of Figure~\ref{fig:K2_networkproduction} displays the total network productivity (i.e., expected success rate of establishing a friendship) achieved by our HGT algorithm across batches (blue), with error bars indicating $\pm 1$ standard deviation (black) based on 100 replications. For reference, the oracle benchmark (orange) represents the fully informed case where the production parameters $\pintb$ and the agents’ latent types $\vZ$ are known. The results show steady improvement in network productivity across batches, with the average increasing from just above $90$ to near the oracle level by the end of the sixth batch. At the same time, variability across replications declines sharply, as the error bars shrink from nearly $15$ to around $3$, suggesting convergence toward oracle-level performance in most of the 100 simulation runs. The right panel shows regret, defined as the difference in expected network output between the HGT algorithm and the oracle benchmark. This plot essentially mirrors the productivity plot in reverse. We present both plots for the stationary environment and use regret as the primary performance metric in the non-stationary setting discussed next, where total network productivity is affected by changes in agent composition, making regret a more consistent metric for comparison.

\subsection{Case of $K = 3$}
% The second simulation setting examines a three-type environment, drawing on a workplace setting inspired by the real estate salesforce application in \cite{xu2024heterogeneous}. 
We consider a pool of $n = 48$ agents (i.e., workers), each belonging to one of three latent types (e.g., weak-, strong-, or moderate-team players), drawn independently from a multinomial distribution with probabilities $(0.4, 0.3, 0.3)$. As in the previous setting, these latent types are fixed once generated in each simulation repetition and treated as unknown parameters. In each batch, these workers are paired into two-person teams to complete $960$ tasks, yielding an average of $40$ tasks per worker. 
% Following \citet{weigel2024supermodular} and \citet{xu2024heterogeneous} which consider fair workload distributions across workers, we also impose a fair workload constraint. 
We impose the agent-level ``workload'' constraint in (\ref{eqn:constraint_workload_individuallevel}) with bounds $[\bound_l, \bound_r] = [35, 45]$.

The team-level success probability matrix $\pintb$ is calibrated to reflect productivity similar to those in \citet{xu2024heterogeneous}, where agents with medium solo performance tend to be strong team players across all types. We preserve this feature while also aligning the matrix with the target of their average productivity level by setting the success probability for a two-worker team to be
\begin{align*}
\pintb
=
\begin{pmatrix}
0.11 & 0.20 & 0.16 \\
0.20 & 0.66 & 0.45 \\
0.16 & 0.45 & 0.37 
\end{pmatrix},
\end{align*}
with rows and columns ordered by type $(\text{weak}, \text{strong}, \text{moderate})$. The symmetry of $\pintb$ implies that the roles of the two workers in a team are interchangeable.

In this case, we also evaluate our algorithm in a non-stationary environment, where workers may leave and new ones may join over time---a common feature in high-turnover industries. In each batch, we randomly select $\turnover$ of the $50$ workers, simulating them being replaced by new hires whose latent types are unknown and drawn from the same population distribution. For each newcomer, the associated $q_i$ values are reset to an uninformative categorical prior with equal probabilities, reflecting the absence of prior knowledge.

\begin{figure}[H]
    \centering
    \textbf{Regret} \par\vspace{2pt}  % Top title with small vertical space
    \hspace{-2pt}\includegraphics[scale=0.61]{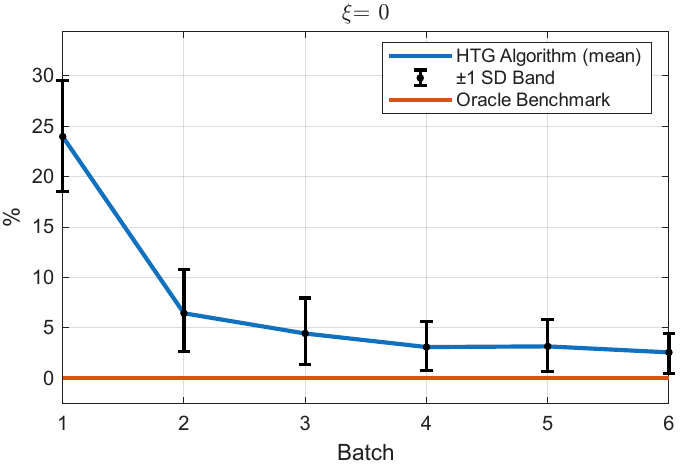}
    \hspace{4pt}\includegraphics[scale=0.61]{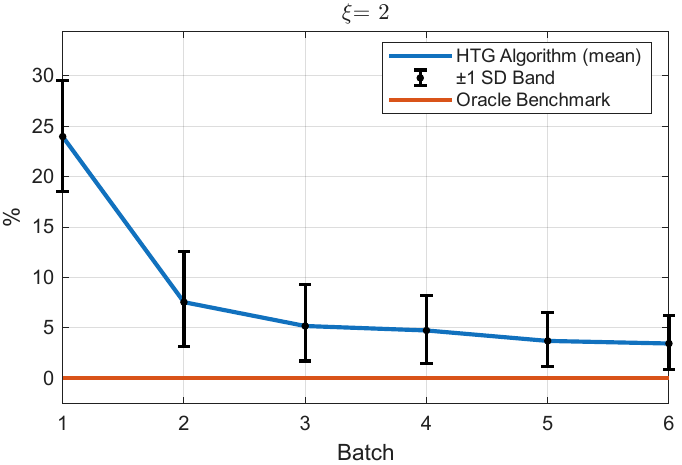}
    
    \vspace{6pt}
    
    \hspace{-2pt}\includegraphics[scale=0.61]{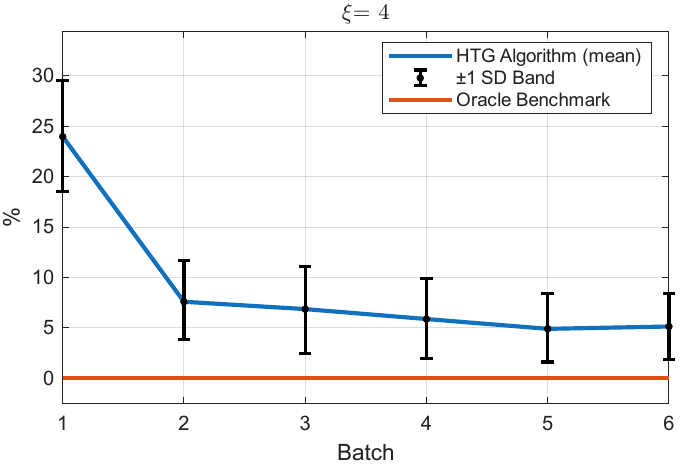}
    \hspace{4pt}\includegraphics[scale=0.61]{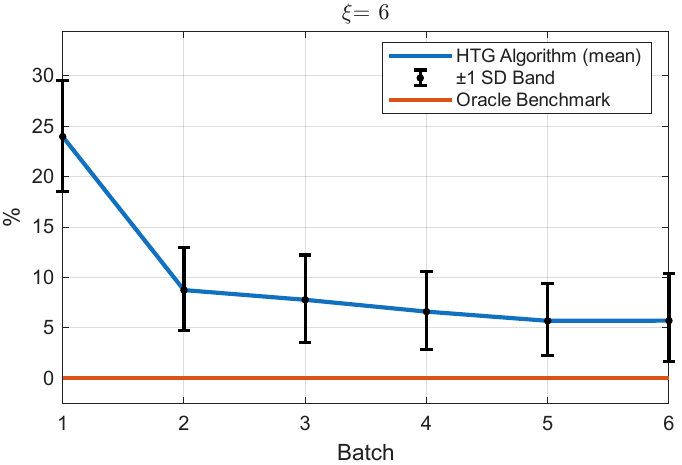}
    
    \caption{\small Average network production performance in \textit{regrets} of the Hybrid Greedy-Thompson (HGT) algorithm (blue), with error bars indicating $\pm 1$ standard deviation (black), under varying levels of worker turnover: $\turnover = 0, 2, 4, 6$. Panels are arranged from top left to bottom right in order of increasing turnover.}
    \label{fig:K3_networkproduction}
\end{figure}

We begin by examining network production performance. Figure~\ref{fig:K3_networkproduction} reports regret curves under four levels of worker turnover ($\turnover = 0, 2, 4, 6$), with panels arranged from top left to bottom right. In each batch, the oracle benchmark is updated in each batch to reflect the current agent composition, and the plotted regret reflects the percentage shortfall in expected successes relative to the oracle. When there is no turnover ($\turnover = 0$), the average regret decreases from about $25$ fewer successful tasks to approximately $2.5$ by batch $6$, relative to the oracle benchmark (which achieves around $290$ expected successes out of $960$ tasks). With $\turnover = 2$ worker replacements per batch, regret still declines, albeit at a slower rate, and converges to a slightly higher level. This pattern becomes more pronounced at higher turnover levels ($\turnover = 4$ and $6$), as shown in the lower panels. This is because the algorithm has no prior knowledge of new hires and must gradually relearn their latent types. Nevertheless, the HGT algorithm continues to yield substantial and consistent gains across all scenarios.

% \begin{figure}[H]
%     \centering
%     \bigskip{}
%     \includegraphics[scale=0.61]{K3_regret_numTurnOver_0.pdf}
%     \includegraphics[scale=0.61]{K3_regret_numTurnOver_2.pdf}
%     \includegraphics[scale=0.61]{K3_regret_numTurnOver_4.pdf}
%     \includegraphics[scale=0.61]{K3_regret_numTurnOver_6.pdf}
%     \medskip{}
%     \caption{\small Average network production performance in \textit{regrets} of the Hybrid Greedy-Thompson (HGT) algorithm (blue), with error bars indicating $\pm 1$ standard deviation (black), under varying levels of worker turnover: $\turnover = 0, 2, 4, 6$. Panels are arranged from top left to bottom right in order of increasing turnover.}
%     \label{fig:K3_networkproduction}
% \end{figure}

Table~\ref{tab:K3_falselabelingrates} reports the false labeling rates, which decrease steadily across batches for each turnover level, indicating improved accuracy in type estimation over time. As turnover increases, the rate of improvement slows, since the algorithm must relearn the latent types of newly arrived workers.
% Indeed, misclassifying workers into incorrect types leads to suboptimal team formation, which in turn affects overall performance, as reflected in the regret plots above.
\begin{table}[H]
\centering
\begin{tabular}{lcccccc}
\toprule
Batch & 1 & 2 & 3 & 4 & 5 & 6 \\
\midrule
$\text{FLR (\%)}$, $\turnover = 0$ & 32.93 & 24.54 & 19.02 & 18.02 & 16.13 & 14.90 \\
$\text{FLR (\%)}$, $\turnover = 2$ & 32.93 & 25.54 & 20.96 & 19.13 & 17.85 & 15.17 \\
$\text{FLR (\%)}$, $\turnover = 4$ & 32.93 & 25.96 & 20.81 & 19.23 & 17.10 & 16.08 \\
$\text{FLR (\%)}$, $\turnover = 6$ & 32.93 & 26.67 & 22.38 & 18.29 & 17.96 & 17.38 \\
\bottomrule
\end{tabular}
\caption{\small False Labeling Rates (FLR) across batches, reported in percentages.}
\label{tab:K3_falselabelingrates}
\end{table}

\begin{figure}[H]
    \centering
    \bigskip{}
    \includegraphics[scale=0.55]{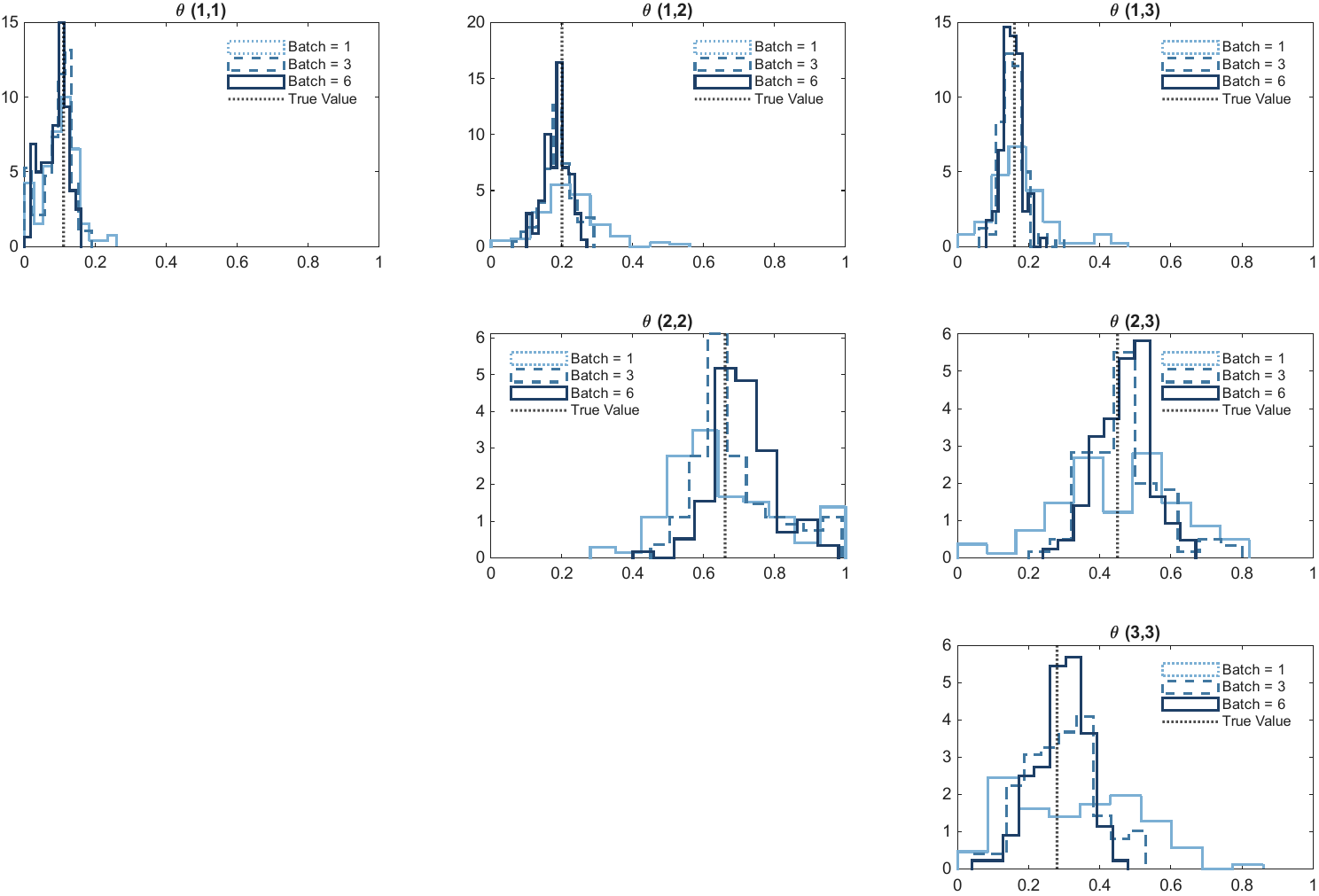}
    \medskip{}
    \caption{\small  Histograms of empirical posteriors for $\pintb$ based on batched variational estimates after batch $1$ (light blue), $3$ (blue), and $6$ (dark blue).}
    \label{fig:K3_empirical_posterior}
\end{figure}

For completeness, we also plot the empirical posterior distributions of each $\pintb$ element for the case of $\turnover = 0$ in Figure~\ref{fig:K3_empirical_posterior}. The histograms confirm that our algorithm progressively sharpens parameter estimates through Bayesian updating. Results under other turnover levels are similar, as agent replacement does not affect the estimation of $\pintb$. We therefore omit those plots.

\section{Conclusion} \label{sec:conclusion}
In this paper, we develop a novel adaptive, or \textit{online}, network formation algorithm that pairs agents, represented as nodes, into edges to maximize total network output, measured by accumulated edge weights. We adopt a weighted stochastic block model, where each agent belongs to one of $K$ latent types that capture unobserved heterogeneity. Only the pairwise outcomes (edge weights), which depend on the types of the paired agents, are observed, while individual node-level contributions remain unobservable.

We formulate the problem of online network formation for improved production as a Batched Multi-Armed Bandit problem, where each type pair corresponds to an arm. In each batch, we assign agents into pairs (i.e., form the network) according to a pre-specified policy. Based on the observed outcomes, we employ a variational approximation to estimate the pairwise expected production parameter and the latent types of agents. These estimates, as justified by our asymptotic analysis, can be treated as Gaussian or categorical signals, and incorporated into updates of the underlying parameters. We then incorporate the updated posteriors into a maximum-weight matching problem to determine the pairing policy for the next batch. Iterating this procedure across batches leads to stochastically improved network-level production.

Our algorithm offers several key advantages. First, the WSBM framework captures heterogeneous complementarities in a nonparametric manner (i.e., without imposing a functional form on the $K \times K$ matrix of expected outcomes), providing greater flexibility than models like the strongest/weakest link, where outcomes depend solely on the better or worse worker. Second, the batched setting enables within-batch asymptotics, allowing us to invoke the multi-stage asymptotic representation theorem of \cite{hirano2023asymptotic}, which offers a much simpler approximation for algorithm design. Third, the batch-wise, adaptively updated policy naturally accommodates nonstationary environments, where production parameters or agent types may evolve over time. Finally, by solving a maximum-weight matching problem, our algorithm can incorporate practical constraints (e.g., to balance workload) that commonly arise in organizational settings. Simulations demonstrate that our approach boosts productivity over a small number of batches while simultaneously improving estimates of both production parameters and latent types.

% \newpage
\bibliographystyle{asa}
\bibliography{references}

% \clearpage
\appendix
\section{Proofs} \label{appendix:proofs}

\subsection{Proof of Proposition~\ref{prop:LAN}}
For notational simplicity, we omit the batch index $t$ in the proof. 

\begin{proof}[Proof of Part i)]
The joint likelihood ratio can be decomposed as follows:
\begin{equation} \label{eqn:proof_likelihooddecomposition}
\begin{aligned}
\lr_{\f}^{(\n,\m)}(\llogizb,\lpintb)
&= \log\frac{\f_{\logizb+\frac{\llogizb}{\sqrt{\n}},\pintb+\frac{\lpintb}{\sqrt{\m}}}(\vY,\vZ)}{\f_{\logizb,\pintb}(\vY,\vZ)}  \\
%&= \log \left(\prod_{i=1}^{\n}\pz_{\logiz+\frac{\llogiz}{\sqrt{\n}}}(Z_i)\right)\p_{\pintb+\frac{\lpintb}{\m}}(\vY|\vZ) - \log \left(\prod_{i=1}^{\n}\pz_{\logiz}(Z_i)\right)\p_{\pintb}(\vY|\vZ)  \\
&= \sum_{i=1}^{\n}\log\frac{\pz_{\logiz+\frac{\llogiz}{\sqrt{\n}}}(\Z_i)}{\pz_{\logiz}(\Z_i)} + \sum_{j=1}^{\m}\log\frac{\p_{\pint + \frac{\lpint}{\sqrt{\m}}}(Y_j|\Zjl,\Zjr)}{\p_{\pint}(Y_j|\Zjl,\Zjr)}  \\
&\defeq \lr^{(\n)}_{\logizb}(\llogizb) + \lr^{(\m)}_{\pintb}(\lpintb).
\end{aligned}
\end{equation}

For the first term, using the fact that $\logiz_{a} = \log\frac{\pz(a)}{1 - \sum_{l=1}^{K-1}\pz(b)}$ implies $\pz_{\logiz}(a) = \frac{e^{\logiz_{a}}}{1+\sum_{b=1}^{K-1}e^{\logiz_{b}}}$ for $a, b = 1,\dots,K-1$, we have
\begin{align*}
\lr^{(\n)}_{\logizb}(\llogizb) 
&= \sum_{i=1}^{\n}\left[\left(\logiz_{\Z_i} + \frac{\llogiz_{\Z_i}}{\sqrt{\n}} - \logiz_{\Z_i}\right) - \log\frac{1+\sum_{b=1}^{K-1}e^{\logiz_{b}+\frac{\llogiz_{b}}{\sqrt{\n}}}}{1+\sum_{b=1}^{K-1}e^{\logiz_{b}}}\right]  \\
&= \sum_{a=1}^{K-1}\sum_{i=1}^{\n}\indicator\{\Z_i = a\}\left[\left(\logiz_{a} + \frac{\llogiz_{a}}{\sqrt{\n}} - \logiz_{a}\right) - \log\frac{1+\sum_{b=1}^{K-1}e^{\logiz_{b}+\frac{\llogiz_{b}}{\sqrt{\n}}}}{1+\sum_{b=1}^{K-1}e^{\logiz_{b}}}\right]  \\
&= \sum_{a=1}^{K-1}\llogiz_{a}\CS^{(\n)}_{\logizb}(a) - \frac{1}{2}\sum_{a=1}^{K-1}\sum_{b=1}^{K-1}\llogiz_{a}\QV_{\logizb}(a,b)\llogiz_{b} + o_P(1), 
\end{align*}
where
\begin{align*}
\CS^{(\n)}_{\logizb}(a) = \frac{\partial\lr_{\logizb}(\llogizb)}{\partial\llogiz_{a}}\bigg|_{\llogiz_{a} = 0} 
&= \frac{1}{\sqrt{\n}}\sum_{i=1}^{\n}\left(\indicator\{\Z_i = a\} - \frac{e^{\logiz_{a}}}{1+\sum_{b=1}^{K-1}e^{\logiz_{b}}}\right) \\
&= \frac{1}{\sqrt{\n}}\sum_{i=1}^{\n}\left(\indicator\{\Z_i = a\} - \pz(a)\right),  
\end{align*}
and
\begin{align*}
\QV_{\logizb}(a,b) = 
      \begin{dcases}
      \frac{\partial^2\lr_{\logizb}(\llogizb)}{\partial\llogiz_{a}\partial\llogiz_{a}}\bigg|_{\llogiz_{a}=0} = \frac{e^{\logiz_{a}}(1+\sum_{b=1}^{K-1}e^{\logiz_{b}}) - e^{\logiz_{a}}e^{\logiz_{a}}}{(1+\sum_{b=1}^{K-1}e^{\logiz_{b}})^2} = \pz(a)(1-\pz(a)), & \text{if $a = b$,} \\
      \frac{\partial^2\lr_{\logizb}(\llogizb)}{\partial\llogiz_{a}\partial\llogiz_{b}}\bigg|_{\llogiz_{a}=\llogiz_{b}=0} 
      = \frac{e^{\logiz_{a}}e^{\logiz_{b}}}{(1+\sum_{b=1}^{K-1}e^{\logiz_{b}})^2} = \pz(a)\pz(b),       & \text{if $a\neq b$.}
      \end{dcases} 
\end{align*}

For the second term, we have
\begin{align*}
\lr^{(\m)}_{\pintb}(\lpintb) 
=&~ \sum_{j=1}^{\m}\log\frac{\p_{\pint + \frac{\lpint}{\sqrt{\m}}}(Y_j|\Zjl,\Zjr)}{\p_{\pint}(Y_j|\Zjl,\Zjr)}  \\
=&~ \sum_{a=1}^{K}\sum_{b=1}^{K}\sum_{j=1}^{\m} \indicator\{\Zjl = a,\Zjr = b\} \log\frac{\p_{\pint_{a b} + \frac{\lpintab}{\sqrt{\m}}}(Y_j|a,b)}{\p_{\pint_{a b}}(Y_j|a,b)}  \\
=&~ \sum_{a=1}^{K}\sum_{b=1}^{K}\Bigg(\frac{\lpintab}{\sqrt{\m}}\sum_{j=1}^{\m} \indicator\{\Zjl = a,\Zjr = b\} \score_{\pint}(Y_j|a,b) \\
&~~~~~~~~~~~~~~ + \frac{1}{2}\frac{\lpintab^2}{\m}\sum_{j=1}^{\m} \indicator\{\Zjl = a,\Zjr = b\}\FI_{\pint}(a,b) + \op(1)\Bigg)  \\
=&~ \sum_{a=1}^{K}\sum_{b=1}^{K}\left(\lpintab\CS^{(\m)}_{\pintb}(a,b) - \frac{1}{2}\lpintab^2\QV_{\pintb}(a,b)\right) + \op(1) 
\end{align*}
where
\begin{align*}
\CS^{(\m)}_{\pintb}(a,b) &= \frac{1}{\sqrt{\m}}\sum_{j=1}^{\m} \indicator\{\Zjl = a,\Zjr = b\} \score_{\pint}(Y_j|a,b), \\
\QV_{\pintb}(a,b) &= \policy(a,b)\FI_{\pint}(a,b).
\end{align*}
The third equality follows directly from the individual LAN condition in Assumption~\ref{assm:LAN_individual}, part (a). The final equality follows from the law of large numbers, together with the fact that $\policy(a,b)$ satisfies the clipping condition (\ref{eqn:constraint_clipping}), implying that $\frac{1}{\m}\sum_{j=1}^{\m} \indicator\{\Zjl = a,\Zjr = b\} \to \policy(a,b)$, for all $a,b=1,\dots,K$.
\end{proof}

\begin{proof}[Proof of Part ii)]
This part follows from a Lindeberg central limit theorem.
\end{proof}

\begin{proof}[Proof of Part iii)]
In the complete model $\f_{\logizb,\pintb}$, the MLE for $\pzb$ is given by
\begin{align*}
\hat\pz(a) = \frac{1}{\n}\sum_{i=1}^{\n}\left(\indicator\{\Z_i = a\}\right)
\end{align*}
and $\hat\logiz_{a} = \log\pz(a) - \log(1 - \sum_{l=1}^{K-1}\pz(l))$ for $a = 1,\dots,K-1$. From standard exponential family theory (see \citet[Lemma~1]{bickel2013asymptotic}), it follows that
\begin{align}  \label{eqn:proof_MLE_f_I}
\sqrt{\n}(\hat\logizb_f^\MLE - \logizb) \sim \mathcal{N}\left(\mfzero, \QVb_{\logizb}^{-1}\right).
\end{align}

Turning next to $\pintb$, Assumption~\ref{assm:LAN_individual}(b) ensures the asymptotic normality on the MLE for each individual production parameter $\pint_{a b}$. Combining with the LAN result established above, the maximum likelihood estimate based on $\f_{\logizb,\pintb}$ satisfies
\begin{equation} \label{eqn:proof_MLE_f_II}
\sqrt{\m}(\hat\pintb_f^\MLE - \pintb) \sim \mathcal{N}\left(\mfzero, \QVb_{\pintb}^{-1}\right).
\end{equation}
See, e.g., Section~7.4 of \citet{van2000asymptotic}.
\end{proof}

\subsection{Proof of Theorem~\ref{thm:fg_equal}}
For notational simplicity, we omit the batch index $t$ in the proof. 

Our proof builds on techniques presented in \cite{bickel2009nonparametric} and \cite{bickel2013asymptotic}. Following their approach, we introduce the notion of \textit{likelihood modularity}, which in our setting is defined as
\begin{align*}
\modula^{(\n,\m)}(\vY,\vZ) = \sup_{\logizb,\pintb}\log\f_{\logizb,\pintb}(\vY,\vZ).
\end{align*}
We assume that the likelihood modularity can be expressed as
\begin{align*}
\modula^{(\n,\m)}(\vY,\ve) = \n\sum_{a=1}^{K}\pz_a(\ve)\log(\pz_a(\ve)) + \mn\modulafun\left(\frac{\vO(\vY,\ve)}{\mn},\frac{\vm(\ve)}{\mn}\right), 
\end{align*}
where
\begin{align*}
& \m_{a b}(\ve) = \sum_{j = 1}^{\m} \indicator\{\e_{i_1(j)} = a, \e_{i_2(j)} = b\}, ~~
\n_{a}(\ve) = \sum_{i = 1}^{\n} \indicator\{\e_i = a\} , \\
& O_{a b}^r(\vY, \ve) = \sum_{j = 1}^{\m}\indicator\{\e_{i_1(j)} = a, \e_{i_2(j)} = b\}\Y_j^r, ~~ \pz_a(\ve) = \frac{\n_{a}(\ve)}{\n},
\end{align*} 
for $\ve\in\{1,\dots,K\}^n$ and appropriate, finite power $r = 1,2,\dots$ of $\Y_j$ used in the analysis, which may vary depending on the distribution of $Y_j$. For instance, $r = 1$ suffices for binary $Y_j$ outcomes modeled as binomial (as in \cite{bickel2013asymptotic} for the classical stochastic block model), and $r = 1, 2$ suffices for continuous $Y_j$ outcomes modeled as Gaussian. The quantity $\vO^r$ collects the $O_{ab}^r$ values for a given $r$, and $\vO$ stacks $\vO^r$ for all relevant powers. Similarly, $\vm$ collects the $\m_{ab}$ terms.

Define the confusion matrix $\vR\in[0,1]^{K\times K}$ as
\begin{align*}
[\vR(\vZ,\ve)](a,b) = \frac{1}{\n}\sum_{i=1}^{\n}\indicator\{\Z_i = b, \e_i = a\}.
\end{align*}
Let $\vmoment_{\pint}^r(a,b) = \Exp_{\pint}(Y_j^r \,|\, \Zjl = a, \Zjr = b)\times\policy(a,b)$ denote the $r$-th moment of an $(a, b)$ pair, weighted by its sampling probability. Let the $K\times K$ matrix $\vmomentb_{\pintb}^r$ collect the values of  $\vmoment_{\pint}^r(a,b)$. For brevity, write $(\vR\vmomentb_{\pintb}^r\vR^\trans)(\ve)$ to denote the product $\vR(\vZ,\ve)\vmomentb_{\pintb}^r\vR(\vZ,\ve)^\trans$, and define
\begin{align*}
\vX^r(\ve) \defeq \mn^{-1}\vO^r(\vY,\ve) - (\vR\vmomentb_{\pintb}^r\vR^\trans)(\ve).
\end{align*}
The conditions that follow formalize the assumptions necessary for our analysis.

\medskip

\begin{assumption} \label{assm_BickelChen2009}
We assume, for each required power $r = 1, 2, \dots$, the following two concentration inequalities hold:
\begin{align} \label{eqn:BickelChen2009_I}
\prob\left(\max_{\ve}\|\vX^r(\ve)\|_{\infty} > \epsilon\right)
\leq
2K^{\n+2}\exp\left(-C_1\epsilon^2\mn\right)
\end{align}
for all $\epsilon \leq c_1$; and
\begin{align} \label{eqn:BickelChen2009_II}
\prob\left(\max_{\ve:|\ve - \vZ|\leq s}\|\vX^r(\ve) - \vX^r(\vZ)\|_{\infty} > \epsilon\right) \leq 2 \begin{pmatrix}\n \\ s\end{pmatrix}K^{s+2}\exp\left(-C_2\frac{\n}{s}\epsilon^2\mn\right)
\end{align}
for all $\epsilon \geq c_2\frac{s}{\n}$. Here, the constants $c_1$, $c_2$, $C_1$, and $C_2$ depends on the production model $\p_{\pint}$ and the moment order $r$.
\end{assumption}

\medskip

\begin{assumption} \label{assm:likelihoodmodularity}
We assume that
\begin{itemize}
\item[(a)] The function $\ve \mapsto \modulafun((\vR\vmomentb_{\pintb}\vR^\trans)(\ve), \vR\policyb\vR^\trans(\ve))$ is maximized by any $\ve\in\mathcal{S}_{\vZ}$.
\item[(b)] $\modulafun$ is Lipschitz continuous in its arguments; The directional derivatives 
\begin{align*}
\frac{\partial^2\modulafun}{\partial\epsilon^2}\left(\vA_0+\epsilon(\vA_1-\vA_0), \vB_0 + \epsilon(\vB_1 - \vB_0) \right)|_{\epsilon = 0^{+}} 
\end{align*}
are continuous in $(\vA_1,\vB_1)$ for all $(\vA_0,\vB_0)$ in a neighborhood of $(\diag(\pz)\vmoment_{\pintb}\diag(\pz),\diag(\pz))$.
\item[(c)] Let $G(\vR,\vmomentb) = \modulafun(\vR\vmomentb_{\pintb}\vR^\trans, \vR\policyb\vR^\trans)$, assume that 
\begin{align*}
\frac{\partial G((1-\epsilon)\diag(\pz) + \epsilon\vR, \vmomentb)}{\partial\epsilon}\bigg|_{\epsilon = 0^{+}} < -C < 0,
\end{align*}
for all $\vR \in \{\vR : \vR\geq 0, \vR^\trans\mfone = \pz\}$.
\end{itemize}
\end{assumption}
\medskip

\begin{proof}[Proof of Part (i)]
We proceed by considering two separate cases: when $\ve$ is far from $\vZ$ and when $\ve$ is close to $\vZ$. To formalize this, define the set
\begin{align*}
E_{\delta_\n} \defeq \left\{\ve: \big|\modulafun((\vR\vmomentb_{\pintb}^r\vR^\trans)(\ve),\vR\policyb\vR^\trans(\ve)) - \modulafun((\vR\vmomentb_{\pintb}^r\vR^\trans)(\vZ),\vR\policyb\vR^\trans(\vZ))\big| \geq \delta_\n \right\}
\end{align*}
where $\delta_\n$ is a sequence converges slowly to zero. 

By (\ref{eqn:BickelChen2009_I}), which implies that $\mn^{-1}\vO^r(\vY,\ve) \to (\vR\vmomentb_{\pintb}^r\vR^\trans)(\ve)$ uniformly over $\ve$, and by the continuity of $\modulafun$, we obtain
\begin{align*}
\prob\left(\max_{\ve}\left|\modulafun\left(\frac{\vO^r(\vY,\ve)}{\mn},\frac{\vm(\ve)}{\mn}\right) - \modulafun((\vR\vmomentb_{\pintb}^r\vR^\trans)(\ve),\vR\policyb\vR^\trans(\ve))\right| \geq \frac{\delta_\n}{2} \right) = o(1).
\end{align*}
Consequently, for all $\ve \in E_{\delta_\n}$, we have 
\begin{align*}
\modulafun\left(\frac{\vO^r(\vY,\ve)}{\mn},\frac{\vm(\ve)}{\mn}\right) \leq \modulafun((\vR\vmomentb_{\pintb}^r\vR^\trans)(\vZ),\vR\policyb\vR^\trans(\vZ)) - \frac{\delta_\n}{2} + o_P(\delta_\n),
\end{align*}
and thus,
\begin{equation} \label{eqn:proof_op1_I}
\begin{aligned} 
 &~ \sum_{\ve\in E_{\delta_\n}} e^{\sup_{\logizb,\pintb}\log\f_{\logizb,\pintb}(\vY,\ve)} \\
= &~ \sum_{\ve\in E_{\delta_\n}} e^{\n\sum_{a=1}^{K}\pz_a(\ve)\log(\pz_a(\ve)) + \mn\modulafun((\vR\vmomentb_{\pintb}^r\vR^\trans)(\ve),\vR\policyb\vR^\trans(\ve))} \\
= &~ \sum_{\ve\in E_{\delta_\n}} e^{O(\n) + \mn(\modulafun((\vR\vmomentb_{\pintb}^r\vR^\trans)(\vZ),\vR\policyb\vR^\trans(\vZ)) - \frac{\delta_\n}{2} + o_P(\delta_\n))} \\
= &~ e^{\mn\modulafun((\vR\vmomentb_{\pintb}^r\vR^\trans)(\vZ),\vR\policyb\vR^\trans(\vZ))} e^{O(\n) - \frac{\mn\delta_\n}{2} + o_P(\mn\delta_\n)}K^\n  \\
= &~ e^{\sup_{\logizb,\pintb}\log\f_{\logizb,\pintb}(\vY,\vZ)}o_P(1) \\
\end{aligned}
\end{equation}
where the final equality follows by choosing $\delta_\n$ to decay slowly enough so that $\mn\delta_\n \gg \n$.

We now consider the case where $\ve \notin E_{\delta_\n}$, meaning that $\ve$ is close to $\vZ$. Let $|\ve - \vZ|$ denote $\sum_{i=1}^{\n}\indicator\{\ve_i \neq \vZ_i\}$, and let $\mathcal{S}(\vZ)$ denote the set of equivalent labels $\vZ^\prime$ up to permutation $\tau:\{1,\dots,K\}\to\{1,\dots,K\}$, that is, $\vZ^\prime = \tau(\vZ)$ for some $\tau$. For a given labeling $\ve$, define $\bar{\ve} = \textrm{arg\,min}_{\tau} |\tau(\ve) - \vZ|$. Applying inequality (\ref{eqn:BickelChen2009_II}), we have
\begin{align*}
 &~ \prob\left( \max_{\ve\notin\mathcal{S}(\vZ)} \|\vX^r(\bar{\ve}) - \vX^r(\vZ)\|_{\infty} > \epsilon\frac{|\bar{\ve} - \vZ|}{\n} \right) \\
\leq &~ \sum_{s = 1}^{n} \prob\left( \max_{\ve:|\bar{\ve} - \vZ| = s} \|\vX^r(\bar{\ve}) - \vX^r(\vZ)\|_{\infty} > \epsilon\frac{s}{\n} \right) \\
\leq &~ \sum_{s = 1}^{n} 2 K^K n^s K^{s+2} \exp\left(-C\mn\frac{s}{\n}\epsilon^2\right) \to 0,
\end{align*}
where the second inequality follows from (\ref{eqn:BickelChen2009_II}). The final convergence to zero holds under the assumption that $\mn \gg \n\log\n$, which allows us to choose $\epsilon\to 0$ such that $\mn\epsilon^2 \gg \n\log\n$. This result shows that $$\max_{\ve\notin\mathcal{S}(\vZ)} \|\vX^r(\bar{\ve}) - \vX^r(\vZ)\|_{\infty} = o_P(|\bar{\ve} - \vZ|/n).$$ As a result, for any $\ve\notin E_{\delta_\n}$, we have
\begin{align*}
\left\| \frac{\vO^r(\vY,\bar{\ve})}{\mn} - \frac{\vO^r(\vY,\vZ)}{\mn} \right\|_{\infty} = \frac{|\bar{\ve} - \vZ|}{\n}o_P(1) + \left\|(\vR\vmomentb_{\pintb}^r\vR^\trans)(\bar{\ve}) - (\vR\vmomentb_{\pintb}^r\vR^\trans)(\vZ)\right\|_{\infty}.
\end{align*}
This further implies
\begin{align} \label{eqn:res_convergence_I}
\frac{\vO^r(\vY,\bar{\ve})}{\mn} - \frac{\vO^r(\vY,\vZ)}{\mn} = \left((\vR\vmomentb_{\pintb}^r\vR^\trans)(\bar{\ve}) - (\vR\vmomentb_{\pintb}^r\vR^\trans)(\vZ)\right)(1 + o_P(1)),
\end{align}
where $o_P(1)$ is uniform over $\ve$.

From Assumption~\ref{assm:likelihoodmodularity}(c), we have
\begin{align*}
\frac{\partial}{\partial\epsilon}\modulafun\left((1-\epsilon)(\vR\vmomentb_{\pintb}^r\vR^\trans)(\vZ) + \epsilon\vA, (1-\epsilon)(\vR\policy\vR^\trans)(\vZ)+\epsilon\vB\right)\bigg|_{\epsilon = 0^{+}} < -\Omega_P(1)
\end{align*}
for all $(\vA,\vB)$ in a neighborhood of $((\vR\vmomentb_{\pintb}^r\vR^\trans)(\vZ), (\vR\policy\vR^\trans)(\vZ))$. Here, $\Omega_P(a_n)$ denotes a quantity that is bounded below by $ca_n$, for some constant $c>0$, with probability approaching one. In view of the result in (\ref{eqn:res_convergence_I}), together with the facts that $\|\vO^r(\vY,\vZ)/\mn - (\vR\vmomentb_{\pintb}^r\vR^\trans)(\vZ)\| = o_P(1)$ and $\|\vm(\vZ)/\mn - (\vR\policy\vR^\trans)(\vZ)\| = o_P(1)$, both of which follow from (\ref{eqn:BickelChen2009_I}), we apply the the directional derivative argument outlined in \citet[page 19]{bickel2013asymptotic} to conclude that
\begin{align*}
\frac{\partial}{\partial\epsilon}\modulafun\left((1-\epsilon)\frac{\vO^r(\vY,\vZ)}{\mn} + \epsilon\vA, (1-\epsilon)\frac{\vm(\vZ)}{\mn}+\epsilon\vB\right)\bigg|_{\epsilon = 0^{+}} < -\Omega_P(1),
\end{align*}
uniformly for $(\vA,\vB)$ in a neighborhood of $(\frac{\vO^r(\vY,\vZ)}{\mn}, \frac{\vm(\vZ)}{\mn})$. 

Now, since $\ve\notin E_{\delta_\n}$ and $\delta_\n \to 0$, we know from the directional derivative argument discussed above that 
\begin{align}
\modulafun\left(\frac{\vO^r(\vY,\ve)}{\mn}, \frac{\vm(\ve)}{\mn}\right) - \modulafun\left(\frac{\vO^r(\vY,\vZ)}{\mn}, \frac{\vm(\vZ)}{\mn}\right) \leq -\frac{1}{\n}\Omega_P(|\bar{\ve} - \vZ|),
\end{align}
where $\Omega_P(|\bar{\ve} - \vZ|)$ is uniform over $\ve$. Therefore, for all $\ve$ such that $|\bar{\ve} - \vZ| = s$, we have
\begin{align*}
 &~ \sup_{\logizb,\pintb}\log\f_{\logizb,\pintb}(\vY,\ve) - \sup_{\logizb,\pintb}\log\f_{\logizb,\pintb}(\vY,\vZ)  \\
\leq &~ \n\left|\sum_{a=1}^{K}\left(\pz_a(\ve)\log(\pz_a(\ve)) - \pz_a(\vZ)\log(\pz_a(\vZ))\right)\right|  \\
&~ + \mn\left(\modulafun\left(\frac{\vO(\vY,\ve)}{\mn},\frac{\vm(\ve)}{\mn}\right) - \modulafun\left(\frac{\vO(\vY,\vZ)}{\mn},\frac{\vm(\vZ)}{\mn}\right)\right)  \\
\leq &~ \n O(1) - \mn\Omega_P(s/n).
\end{align*}
This leads to the bound
\begin{equation} \label{eqn:proof_op1_II}
\begin{aligned}
 &~ \sum_{\ve\notin E_{\delta_\n}, \ve\notin\mathcal{S}(\vZ)} e^{\sup_{\logizb,\pintb}\log\f_{\logizb,\pintb}(\vY,\ve)} \\
= &~ \sum_{s=1}^{\n}\sum_{\ve:|\bar{\ve} - \vZ| = s} e^{\sup_{\logizb,\pintb}\log\f_{\logizb,\pintb}(\vY,\ve)} \\
= &~ \sum_{s=1}^{\n}\sum_{\ve:|\bar{\ve} - \vZ| = s} e^{\sup_{\logizb,\pintb}\log\f_{\logizb,\pintb}(\vY,\vZ)}e^{\sup_{\logizb,\pintb}\log\f_{\logizb,\pintb}(\vY,\ve) - \sup_{\logizb,\pintb}\log\f_{\logizb,\pintb}(\vY,\vZ)} \\
\leq &~ \sum_{s=1}^{\n}\sum_{\ve:|\bar{\ve} - \vZ| = s} e^{\sup_{\logizb,\pintb}\log\f_{\logizb,\pintb}(\vY,\vZ)}e^{O(\n) - \mn\Omega_P(s/n)} \\
\leq &~ \sum_{s=1}^{\n} e^{\sup_{\logizb,\pintb}\log\f_{\logizb,\pintb}(\vY,\vZ)}K^Kn^sK^se^{O(\n) - \mn\Omega_P(s/n)} \\
\leq &~ \sum_{s=1}^{\n} e^{\sup_{\logizb,\pintb}\log\f_{\logizb,\pintb}(\vY,\vZ)}K^Ke^{s(log\n + \log K + O(\n) - \Omega_P(\mn/n))} \\
= &~ e^{\sup_{\logizb,\pintb}\log\f_{\logizb,\pintb}(\vY,\vZ)}o_P(1).
\end{aligned}
\end{equation}
Combining this with the contribution from $\ve\in E_{\delta_n}$ ((as controlled in Eq.(\ref{eqn:proof_op1_I})), we conclude that for all $(\logizb, \pintb)$,
\begin{align} \label{eqn:proof_op1_III}
\sum_{\ve\notin\mathcal{S}(\vZ)} \f_{\logizb,\pintb}(\vY,\ve) \leq \sum_{\ve\notin\mathcal{S}(\vZ)} \sup_{\logizb,\pintb}\f_{\logizb,\pintb}(\vY,\ve) = \sup_{\logizb,\pintb}\f_{\logizb,\pintb}(\vY,\vZ)o_P(1).
\end{align}

Recalling that $\g_{\pzb,\pintb}(\vY) = \sum_{\ve\in\{1,\dots,K\}^{\n}}\f_{\pzb,\pintb}(\vY,\ve)$, we can write, for all $(\logizb, \pintb)$,
\begin{equation} \label{eqn:proof_op1_IV}
\begin{aligned} 
\f_{\pzb,\pintb}(\vY,\vZ) 
\leq \g_{\pzb,\pintb}(\vY) 
=&~ \f_{\pzb,\pintb}(\vY,\vZ) + \sum_{\ve\notin\mathcal{S}(\vZ)}\f_{\pzb,\pintb}(\vY,\ve) \\
\leq&~ \f_{\pzb,\pintb}(\vY,\vZ) + \sup_{\logizb,\pintb}\f_{\logizb,\pintb}(\vY,\vZ)o_P(1) \\
=&~ \f_{\pzb,\pintb}(\vY,\vZ) + o_P(1),
\end{aligned}
\end{equation}
where the first inequality holds since all terms in the sum are nonnegative (i.e., $\f_{\pzb,\pintb}(\vY,\ve) \geq 0$ for any $\ve$), and the second inequality is from (\ref{eqn:proof_op1_III}).

\end{proof}

\begin{proof}[Proof of Part (ii)]
Since the MLE $(\hat\logizb_g^\MLE,\hat\pintb_g^\MLE)$ is obtained by maximizing $\g_{\logizb,\pintb}(\vY)$, it follows from Part (i) that there exists an element $(\hat\logizb_g^\prime,\hat\pintb_g^\prime) \in \mathcal{S}(\hat\logizb_g^\MLE,\hat\pintb_g^\MLE)$ such that
\begin{align} \label{eqn:proof_thm1part2_I}
\left|\f_{\pzb_g^\prime,\pintb_g^\prime}(\vY,\vZ) - \f_{\hat\pzb_f^\MLE,\hat\pintb_f^\MLE}(\vY,\vZ)\right| = o_P(1),
\end{align}
where $(\hat\pzb_f^\MLE,\hat\pintb_f^\MLE)$ is the MLE that maximizes $\f_{\pzb,\pintb}(\vY,\vZ)$. By the LAN result established in Proposition~\ref{prop:LAN} for the complete model $\f$, we then have
\begin{align} \label{eqn:proof_thm1part2_II}
\frac{1}{\sqrt{\n}}\left|\hat\logizb_g^\prime - \hat\logizb_f^\MLE\right| = o_P(1) \textrm{~~and~~} \frac{1}{\sqrt{\m}}\left|\hat\pintb_g^\prime - \hat\pintb_f^\MLE\right| = o_P(1). 
\end{align}
This can be seen more clearly by contradiction: if either equality in (\ref{eqn:proof_thm1part2_II}) does not hold, then (\ref{eqn:proof_thm1part2_I}) is violated.

Combining (\ref{eqn:proof_MLE_f_I}), (\ref{eqn:proof_MLE_f_II}) with (\ref{eqn:proof_thm1part2_II}) completes the proof.
\end{proof}

\subsection{Proof of Theorem~\ref{thm:variational_LAN}}
For notational simplicity, we omit the batch index $t$ in the proof. 

\begin{proof}[Proof of Part (i)]
Recall from (\ref{eqn:elbo}) that he evidence lower bound (ELBO) is given by
\begin{align*} 
\elbo_{\pzb,\pintb;\vq}(\vY) 
=&~ \sum_{i=1}^{\n}\sum_{a=1}^{K}q_{i}(a)\left(-\log\q_{i}(a) + \log\pi(a)\right) \\
&~ + \sum_{j=1}^{\m}\sum_{a=1}^{K}\sum_{b=1}^{K} \qjl(a)\qjr(b)\log \p_{\pint}(Y_j|a,b).
\end{align*}
It is straightforward to verify that $\elbo_{\pzb,\pintb;\vq}(\vY) = \log\f_{\pzb,\pintb}(\vY,\vZ)$ when $\q_{i}(a) = \indicator\{\Z_i = a\}$ for all $i = 1,\dots,\n$ and $a=1,\dots,K$. Hence, letting $$\hat\vq = \hat\vq(\pzb,\pintb) \defeq \argmax_{\vq\in\SQ} \elbo_{\pzb,\pintb;\vq}(\vY),$$ we have $\log\f_{\pzb,\pintb}(\vY,\vZ) \leq \elbo_{\pzb,\pintb;\hat\vq}(\vY)$. 

On the other hand, by (\ref{eqn:KL2}) and the non-negativity of the Kullback–Leibler divergence, the ELBO is always a lower bound on the log marginal likelihood, i.e., $\elbo_{\pzb,\pintb;\vq}(\vY) \leq \log\g_{\pzb,\pintb}(\vY)$ for all $\q$. Combining these inequalities yields
\begin{align} \label{eqn:proof_J_inequalities}
\log\f_{\pzb,\pintb}(\vY,\vZ) \leq \elbo_{\pzb,\pintb;\hat\vq}(\vY)  \leq \log\g_{\pzb,\pintb}(\vY). 
\end{align}

Next, observe that
\begin{align*}
\log\g_{\pzb,\pintb}(\vY) 
=&~ \log\left(\f_{\pzb,\pintb}(\vY,\vZ) + \sum_{\ve\notin\mathcal{S}(\vZ)}\f_{\pzb,\pintb}(\vY,\ve)\right)  \\
=&~ \log\f_{\pzb,\pintb}(\vY,\vZ) + \log\left(1 + \frac{\sum_{\ve\notin\mathcal{S}(\vZ)}\f_{\pzb,\pintb}(\vY,\ve)}{\f_{\pzb,\pintb}(\vY,\vZ)}\right)  \\
=&~ \log\f_{\pzb,\pintb}(\vY,\vZ) + \log\left(1 + \frac{\f_{\pzb,\pintb}(\vY,\vZ)o_P(1)}{\f_{\pzb,\pintb}(\vY,\vZ)}\right)  \\
=&~ \log\f_{\pzb,\pintb}(\vY,\vZ) + o_P(1),
\end{align*}
where the third equality follows from (\ref{eqn:proof_op1_IV}) in the proof of Theorem~\ref{thm:fg_equal}. 

Combining this with (\ref{eqn:proof_J_inequalities}), we conclude that
\begin{align*}
\elbo_{\pzb,\pintb;\hat\vq}(\vY) = \log\f_{\pzb,\pintb}(\vY,\vZ) + o_P(1),
\end{align*}
which completes the proof.
\end{proof}

\begin{proof}[Proof of Part (ii)]
The proof follows using arguments similar to Theorem~\ref{thm:fg_equal}(ii).
\end{proof}

\end{document}